\documentclass[10pt, aps, pra, superscriptaddress, nofootinbib]{revtex4-2}

\usepackage{geometry}
\usepackage{xcolor}
\usepackage{textcomp}
\usepackage{bm}
\usepackage{amsmath}
\usepackage{amssymb}
\usepackage{graphicx}
\usepackage{braket}
\usepackage{upgreek}
\usepackage{soul}
\usepackage[normalem]{ulem}
\usepackage[unicode=true, pdfusetitle, bookmarks=true, 
            bookmarksnumbered=false, bookmarksopen=true,
            bookmarksopenlevel=1, breaklinks=false,
            pdfborder={0 0 1}, backref=false,
            colorlinks=true]{hyperref}
\hypersetup{pdfborderstyle=, linkcolor=blue, 
            urlcolor=blue, citecolor=blue}
\usepackage{orcidlink}

\newcommand{\bi}{\mathbf}

\newcommand{\be}{\begin{equation}}
\newcommand{\ee}{\end{equation}}
\newcommand{\beq}{\begin{eqnarray}}
\newcommand{\eeq}{\end{eqnarray}}

\begin{document}

\title{Dynamical generation and transfer of nonclassical states in strongly interacting light-matter systems in cavities}

\author{Ilia Tutunnikov\,\orcidlink{0000-0002-8291-7335}}
\affiliation{ITAMP, Center for Astrophysics $|$ Harvard \& Smithsonian, Cambridge,
Massachusetts 02138, USA}

\author{Vasil Rokaj\,\orcidlink{0000-0002-0627-7292}}
\email{vasil.rokaj@villanova.edu}
\affiliation{ITAMP, Center for Astrophysics $|$ Harvard \& Smithsonian, Cambridge,
Massachusetts 02138, USA}
\affiliation{Department of Physics, Villanova University, Villanova, Pennsylvania 19085, USA \looseness=-1}
\affiliation{Department of Physics, Harvard University, Cambridge, Massachusetts 02138, USA}

\author{Jianshu Cao\,\orcidlink{0000-0001-7616-7809}}
\affiliation{Department of Chemistry, Massachusetts Institute of Technology, Cambridge,
Massachusetts 02139, USA}

\author{H. R. Sadeghpour\,\orcidlink{0000-0001-5707-8675}}
\affiliation{ITAMP, Center for Astrophysics $|$ Harvard \& Smithsonian, Cambridge,
Massachusetts 02138, USA}

\begin{abstract}
We propose leveraging strong and ultrastrong light-matter coupling to efficiently
generate and exchange nonclassical light and quantum matter states. Two
initial conditions are considered: (a) a displaced quadrature-squeezed
matter state, and (b) a coherent state in a cavity. In both scenarios, polaritons mediate the dynamical generation and transfer of nonclassical states between light and matter. By monitoring the dynamics of both subsystems, we uncover the emergence of beatings in the collective matter oscillations. The beating period depends on the particle density through the vacuum Rabi splitting and peaks sharply under light-matter resonance conditions. For initial condition (a), nonclassicality is efficiently transferred from matter to photons under strong and ultrastrong coupling. However, for initial condition (b), nonclassical photonic states are generated only in the ultrastrong coupling regime due to the counter-rotating terms, highlighting the advantages of ultrastrong coupling. Furthermore, in the ultrastrong coupling regime, distinctive asymmetries relative to cavity detuning emerge in dynamical observables of both light and matter. The nonclassical photons can be extracted through a semi-transparent cavity mirror, while nonclassical matter states can be detected via time-resolved spectroscopy. This work highlights that hybrid polariton states can be utilized for dynamically generating nonclassical states, with potential applications in quantum state transfer.
\end{abstract}

\date{\today}

\maketitle

\section{Introduction}

Strong light-matter coupling is essential for nascent quantum technologies,
enabling efficient and reversible quantum state transfer. Polaritons,
the light-matter quasiparticles, are often best manipulated in cavities~
\cite{Carusotto2013, FriskKockum2019, UltrastrongReview2019, Basov2021},
and have been widely studied in condensed matter~\cite{Appugliese2022, EnkerPRX, 
Hagenmuller2010, Scalari2012, Bayer2017, Ravets2018, Sentef2018, Li2018, Schlawin2019, 
Curtis2019, RokajTopo2023, Schlawin2022}, cold atoms~ \cite{Kasprzak2006, Dudin12, 
Peyronel12, FarokhMivehvar2021}, and molecular systems~\cite{Hutchison2012, Ebbesen2016, 
Jino2016, Yang2021, WuCerrilloCao, Sidler2022, Feist2018, Hou2020, Keeling2020, 
GarciaVidal2021, Balasubrahmaniyam2023}. 
For example, polariton-based systems demonstrated modifications of chemical reactivity
~\cite{Hutchison2012, Hutchison2013, Ebbesen2016, GarciaVidal2021, Sun2024, Xiang2024},
transient diffusivity and transport~\cite{Freixanet2000, Orgiu2015, Hou2020, Pandya2021, Engelhardt2022, *Engelhardt2023,
Sokolovskii2023, Balasubrahmaniyam2023, AroeiraRibeiro2024, ZhouNitzan2024, Sandik2024}, as well as exciton-polariton condensation~\cite{Kasprzak2006, Foerg2019, Latini2019, Keeling2020}. 

Equally important is the ability to generate and manipulate photons in such hybrid 
polaritonic systems. Photon emission statistics has been explored in experiments with 
ultracold atomic gases, leveraging strong Rydberg correlations for deterministic single-photon 
generation~\cite{Dudin12, Peyronel12}. Second-order photon correlation measurements 
confirm the quantum nature of emitted light, achieving high fidelity with high-finesse 
optical cavities~\cite{Kuhn02}. Squeezed light~\cite{Andersen2016} has been demonstrated 
in various optomechanical systems~\cite{LawMirror1995, AspelmeyerReview2014}, e.g., mechanical resonators~\cite{SafaviNaeini2013, Purdy2013}, and ultracold 
atoms~\cite{Brooks2012}.

Here, we propose dynamically generating and transferring nonclassical
states of quantum matter and light by leveraging strong light-matter coupling
in a single-mode cavity~\cite{Takahashi2020, Vasilis2023}. The model
used to describe the composite-system dynamics is the archetypical
Hopfield Hamiltonian (HH)~\cite{Hopfield}, first introduced for studying 
excitonic response in dielectric crystals~\cite{Hopfield}. We show that the
HH applies to various physical systems in cavities, including cold 
harmonically trapped ions~\cite{Vasilis2023}, two-dimensional (2D) electron gases in a strong magnetic field (Landau levels)~\cite{Hagenmuller2010, 
RokajTopo2023, Appugliese2022, EnkerPRX, Scalari2012, Li2018}, and oriented 
polar molecules~\cite{rokaj2017, ParticlesBook2004}.

Trapped ions are ideal for studying quantum phenomena, and ion-cavity
systems have enabled single-photon emission, ion-photon, and ion-ion
entanglement \cite{Kreuter2004, Stute2012, Takahashi2017, Keller2022}.
The position of a trapped ion can be precisely controlled within the cavity~\cite{Walther01}
and Fock, coherent, and squeezed motional states of a harmonically
trapped ion have been generated~\cite{Wineland96}. Coherent transfer
between an ion's internal degrees of freedom and an optical cavity
mode has been demonstrated in several experiments \cite{blatt2013, Kimble2007}.
In contrast, cavity coupling with the trapped ion motional states
is technically more challenging due to the relatively low characteristic
motional frequencies, often in the MHz range~\cite{Keller2022, Hite2013}.

Importantly, it has been shown that Landau levels exhibit ultrastrong coupling to the cavity field
\cite{Scalari2012, Smolka2014, Bayer2017, Ravets2018, Li2018, 
ParaviciniBagliani2019, Keller2020, UltrastrongReview2019} and Landau polariton states have been observed \cite{Hagenmuller2010, Scalari2012, Li2018, Vasilis2019, Keller2020, RokajThesis2022}.
More recently, mechanisms for cavity-induced modification of the integer
Hall effect were proposed \cite{Ciuti2021, Rokaj2022}, and the breakdown
of topological protection has been experimentally demonstrated~\cite{Appugliese2022}.

Further, recent experiments provide evidence that chemical reactions may be substantially affected by strong coupling between molecular vibrational modes and the electromagnetic modes of infrared (IR) microcavities--a phenomenon known as vibrational strong coupling (VSC)~\cite{Thomas2016, Lather2019, Thomas2019, GarciaVidal2021}.
This has led to extensive theoretical work \cite{Li2020, Yang2021, 
AroeiraRibeiro2024, JoelReview, HuoReview2023, horak2024, JoelPolCond} in 
elucidating the mechanism behind the observed effects.
Typically, the rotational degrees of freedom introduce additional
challenges for manipulating molecules. Nevertheless, there is rich
literature on the coherent control of molecules, including techniques
tailored to targeting vibrations \cite{Banin1994, ShapiroBrumer2011}.

This work focuses on the collective dynamics in cavity quantum electrodynamics (QED) systems by 
considering two schemes to generate and transfer nonclassical states
of light and matter in a cavity. In \textbf{Scheme I}, the matter state is
displaced and quadrature-squeezed. The dynamics of both subsystems are monitored, and we uncover the emergence of beatings in the collective matter oscillations. The beating period depends on the detuning and peaks sharply around the light-matter resonance. This phenomenon highlights that collective dynamics of matter can be significantly modified via strong and resonant light-matter interaction in a cavity, even without an external field, i.e., in vacuum. Similar resonant modifications have been observed in chemical reactions in molecules under VSC~\cite{Hutchison2012, Hutchison2013, Ebbesen2016, GarciaVidal2021, Sun2024, Xiang2024}. Looking into the photons, the cavity coupling \emph{transfers}
the quantum state of matter to light, resulting in sub-Poissonian
photon number distribution (PND), i.e., quantum photonic states with no 
classical analogue~\cite{KnightBook2012}. In \textbf{Scheme
II}, the cavity is initiated in a coherent state~\cite{BruneHaroche1996},
which partially moves from one subsystem to another over time. Simultaneously,
the counter-rotating light-matter interaction terms (counter-rotating terms, CRT) in the HH
(see below) induce nonclassical features in light or matter subsystems
depending on the initial shift of the coherent state. In the second scheme, 
the nonclassical features arise entirely from the CRT, underscoring the 
importance of the ultrastrong coupling~\cite{kockum2019ultrastrong, UltrastrongReview2019}. 

Notably, the CRT induce asymmetries in dynamical observables associated with the matter and light subsystems. These asymmetries emerge in the ultrastrong coupling regime when the cavity frequency is scanned around the resonance and are consequences of the asymmetry in the polaritonic branches. In contrast to linear spectral measurements, which may be insensitive to the quantum nature of the system, time-dependent observables involving higher-order moments of the matter or light distributions (e.g., photon number variance) serve as quantum measures sensitive to ultrastrong coupling effects. We provide analytical expressions for the average photon number in both schemes, allowing for a comprehensive description of photon generation over the entire parameter space, including the system parameters and initial state conditions.

We quantify the temporal evolution of nonclassical features in matter and light states using the time-averaged Mandel $Q$ parameter for light and
matter. For Scheme I, we demonstrate that outcoupling the cavity field
through a semi-transparent cavity mirror allows for extracting the nonclassical
photons~\cite{Kimble1997}. In Scheme II applied to trapped ions, for
instance, sideband spectroscopy of motional state population allows
for the measurement of matter quantum state statistics~\cite{Hite2013}. 
Hybrid, polaritonic quantum systems can be utilized for the efficient transfer of nonclassicality from matter to light. Nonclassical states are necessary for sensing beyond the standard quantum limit~\cite{QSensing, QJumps, Catsensor, XieQuantumLimit}, and likely for quantum information technologies~\cite{Gao2010, photoncluster2024}. Our work bridges strong and ultrastrong light-matter coupled systems with quantum science and suggests novel pathways for polaritonic quantum technologies.

\section{Hopfield Hamiltonian Across Different Cavity QED Systems \label{sec:Hopfield-Hamiltonian}}

In recent years, the HH has attracted significant attention due to its
application in several cavity QED systems operating in the ultrastrong
light-matter coupling regime~\cite{kockum2019ultrastrong, UltrastrongReview2019}.
This section explicitly demonstrates how the HH emerges in three example systems in a cavity: cold trapped ions, 2D electron gas in a strong magnetic field, and oriented molecules with harmonic internuclear potential. Thus, in principle, the nonclassical dynamical phenomena discussed hereafter are within experimental reach across various physical systems.
\begin{figure*}[t]
\begin{centering}
\includegraphics{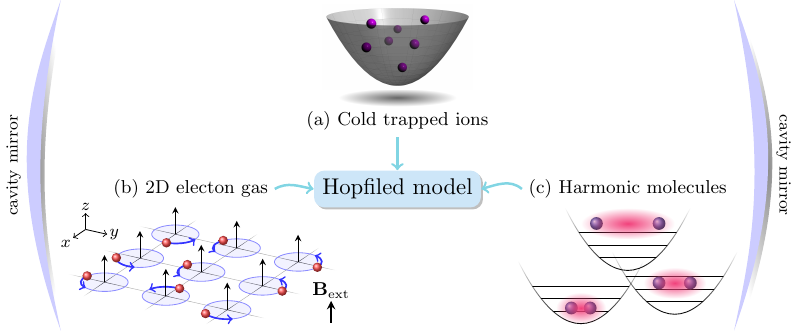}
\end{centering}
\caption{Schematic illustration of the collective coupling to
a homogeneous quantum cavity field in different physical systems, 
including (a) harmonically trapped cold ions, (b) 2D electron gas 
in a strong magnetic field, $\mathbf{B}_\text{ext}$, and 
(c) molecules with harmonic internuclear potential. In all cases, 
the system Hamiltonian can be mapped onto the HH, see Eqs.\;\eqref{eq:ionHam},\;\eqref{eq:2degHopfield},\;and\;
\eqref{eq:molecular-Hamiltonian}. Thus, the dynamical phenomena 
described hereafter can be, in principle, realized in various 
cavity QED systems.}
\label{fig:FIG1} 
\end{figure*}

\subsection{Cold Trapped Ions}

We start by considering a system of $N$ interacting ions confined in a
harmonic potential, coupled to a single-mode quantized cavity field 
[Fig.\;\ref{fig:FIG1}(a)].
Recently, this system was used to study the collective phenomena in the polaritonic ground state~\cite{Vasilis2023}. Here, we focus on the dynamics 
mediated by the coupling between the ions' motion
and the cavity field. In the non-relativistic limit, this system is described 
by the Pauli-Fierz Hamiltonian in the Coulomb gauge, also known as the 
minimal-coupling Hamiltonian\;\cite{ParticlesBook2004, CohenTBook1997}
\begin{equation}
\hat{H}_{\mathrm{ion}}=\frac{1}{2m}\sum\limits _{i=1}^{N}(\mathrm{i}\hbar\mathbf{\nabla}_{i}+g_{0}\hat{\mathbf{A}})^{2}+\sum\limits _{i<l}^{N}W(\hat{\mathbf{r}}_{i}-\hat{\mathbf{r}}_{l})+\sum_{i=1}^{N}\frac{m\Omega^{2}}{2}\hat{\mathbf{r}}_{i}^{2}+\sum\limits _{\nu=x,y}\hbar\omega\left(\hat{a}_{\nu}^{\dagger}\hat{a}_{\nu}+\frac{1}{2}\right),
\label{eq:Pauli-Fierz}
\end{equation}
where $g_{0}$ is the single-particle coupling parameter to the cavity
field (in units of the elementary charge, $e$), $m$
is the mass of the particles, and $\Omega$ is the frequency of the
harmonic trap. The quantized vector potential $\hat{\mathbf{A}}$
in the long-wavelength limit (homogeneous approximation) 
is given by~\cite{ParticlesBook2004, rokaj2017, CohenTBook1997}
\begin{equation}
\hat{\mathbf{A}}=\sum_{\nu=x,y}\sqrt{\frac{\hbar}{2 \epsilon_0\mathcal{V} \omega}}\mathbf{e}_\nu(\hat{a}_{\nu}+\hat{a}_{\nu}^{\dagger}),
\label{eq:AinCoulomb}
\end{equation}
where $\omega=c|k_{z}|$ is the frequency in the quantization volume
$\mathcal{V}$, with wavevector in the $z$ direction, $\epsilon_0$
is the vacuum permittivity, and $\nu=x,y$ denote the two transversal
polarization directions \cite{ParticlesBook2004, CohenTBook1997}.
The operators $\hat{a}_\nu$ and $\hat{a}_\nu^\dagger$ are
the annihilation and creation operators of the photon field obeying
$[\hat{a}_\nu,\hat{a}_{\nu^\prime}^\dagger]=\delta_{\nu\nu^\prime}$.

Expanding the covariant kinetic energy shows that
the homogeneous photon field couples to the total momentum of the
particles, $\hat{\mathbf{A}}\cdot\sum_{i=1}^{N}\nabla_{i}$, implying 
\emph{collective} coupling to the cavity field through the particles' center of 
mass (CM). Thus, we change variables to the CM position $\hat{\mathbf{R}}=N^{-1/2}\sum_{i=1}^{N}\hat{\mathbf{r}}_{i}$,
and the relative positions 
$\hat{\tilde{\mathbf{r}}}_{j}=N^{-1/2}(\hat{\mathbf{r}}_{1}-\hat{\mathbf{r}}_{j})$
($j>1$). The prefactor $N^{-1/2}$ is introduced for mathematical convenience 
like in~\cite{BuschSol}. The operators $\hat{\mathbf{R}}$ and $\hat{\tilde{\mathbf{r}}}_{j}$, 
along with their corresponding momenta commute.
This confirms the independence of CM and relative coordinates 
(App. \ref{sec:APP-ionsApp}).

The scalar trapping potential separates into two
parts: the first one depends on the CM coordinate, while the second depends on the
relative coordinates, without cross-terms between the two, $\sum_{i=1}^{N}\hat{\mathbf{r}}_{i}^{2}=\hat{\mathbf{R}}^{2}+N\sum_{j=2}^{N}\hat{\tilde{\mathbf{r}}}_{j}^{2}-(\sum_{j=2}^{N}\hat{\tilde{\mathbf{r}}}_{j})^{2}$.
The two-body interaction $W\left(\hat{\mathbf{r}}_{i}-\hat{\mathbf{r}}_{l}\right)$
depends only on the relative distances, so it does not affect
the cavity-induced CM motion. The Hamiltonian, therefore, splits into two parts:
(i) $\hat{H}_{\textrm{ion-cm}}$, describing the the CM coupling
to the quantized field $\hat{\mathbf{A}}$, and (ii) $\hat{H}_{\textrm{ion-rel}}$, describing the dynamics of the relative coordinates, decoupled from both $\hat{\mathbf{A}}$ and the CM. Then,
\begin{eqnarray}
\hat{H}_{\textrm{ion-cm}}=\frac{1}{2m}(\textrm{i}\hbar\nabla_{\mathbf{R}}+g_{0}\sqrt{N}\hat{\mathbf{A}})^{2}+\frac{m\Omega^{2}}{2}\hat{\mathbf{R}}^{2}+\sum_{\nu=x,y}\hbar\omega\left(\hat{a}_{\nu}^{\dagger}\hat{a}_{\nu}+\frac{1}{2}\right),
\end{eqnarray}
and $\hat{H}_{\textrm{ion-rel}}$ is presented in App.~\ref{sec:APP-ionsApp}.

We now focus exclusively on the CM part to describe the cavity-matter
dynamics. Since the polarization vectors of the photon field lie in
the $(x,y)$ plane, the $z$ direction is trivial. The light-matter
Hamiltonian then becomes a system of interacting harmonic oscillators,
and importantly, the $x$ and $y$ directions are decoupled,
\begin{equation}
\hat{H}_{\text{ion-cm}}=\sum_{\nu=x,y}\Big[-\frac{\hbar^{2}}{2m}\partial_{R_{\nu}}^{2}+\frac{\textrm{i}g_{0}\hbar}{m}\sqrt{N}\hat{A}_{\nu}\partial_{R_{\nu}}+\frac{m\Omega^{2}}{2}\hat{R}_{\nu}^{2}+\frac{Ng_{0}^{2}}{2m}\hat{A}_{\nu}^{2}+\hbar\omega\left(\hat{a}_{\nu}^{\dagger}\hat{a}_{\nu}+\frac{1}{2}\right)\Big].
\end{equation}
To avoid any confusion, note that $\hat{\mathbf{R}}=(\hat{R}_{x},\hat{R}_{y})=(\hat{X},\hat{Y})$,
$\nabla_{\hat{\mathbf{R}}}=(\partial_{R_{x}},\partial_{R_{y}})=(\partial_{X},\partial_{Y})$
and $\hat{\mathbf{A}}=(\hat{A}_{x},\hat{A}_{y})$. Consequently, without
loss of generality, we can focus only on the $X$ component of the
Hamiltonian. We also suppress the $x$ index in the photon operators for simplicity, i.e., $\hat{a}\equiv \hat{a}_x$.
Finally, to simplify the Hamiltonian, we introduce bosonic operators for matter $\hat{b}=\sqrt{m\Omega/(2\hbar)}[\hat{X}+\hbar/(m\Omega)\partial_X]$. The CM Hamiltonian turns into the well-known HH
\cite{Hopfield, kockum2019ultrastrong} 
\begin{equation}
\hat{H}_{\text{ion-cm}}\!=\!\hbar\Omega\left(\hat{b}^{\dagger}\hat{b}\!+\!\frac{1}{2}\right)+\mathrm{i}\hbar\sqrt{\frac{\Omega g_{0}^{2}N}{4m \epsilon_{0}\mathcal{V}\omega}}\left(\hat{a}\!+\!\hat{a}^{\dagger}\right)(\hat{b}-\hat{b}^{\dagger})+\frac{\hbar Ng_{0}^{2}}{4m\epsilon_{0}\mathcal{V}\omega}\left(\hat{a}\!+\!\hat{a}^{\dagger}\right)^{2}\!+\hbar\omega\left(\hat{a}^{\dagger}\hat{a}\!+\!\frac{1}{2}\right)\!.
\label{eq:ionHam}
\end{equation}

\subsection{2D Electron Gas}

Next, we consider 2D electron gas (2deg) subject to a homogeneous magnetic field perpendicular to the plane. The electrons are coupled to a homogeneous single-mode cavity field [Fig.\;\ref{fig:FIG1}(b)]. 
Landau-level systems in a cavity have been theoretically studied~\cite{Hagenmuller2010, RokajTopo2023}, and, notably, have been experimentally realized. Many interesting phenomena
have been observed, including Landau polariton quasiparticles and modifications of quantum Hall transport~\cite{Appugliese2022, EnkerPRX, Scalari2012, Li2018}.
The Hamiltonian for the quantum Hall system in the cavity is given by
\begin{equation}
\hat{H}_{{\rm {2deg}}}=\frac{1}{2m}\sum_{i=1}^{N}(\hat{\bm{\pi}}_{i}+e\hat{\mathbf{A}})^{2}+\sum\limits _{i<l}^{N}W(\hat{\mathbf{r}}_{i}-\hat{\mathbf{r}}_{l})+\hbar\omega\left(\hat{a}^{\dagger}\hat{a}+\frac{1}{2}\right),
\label{eq:2degH}
\end{equation}
where $\hat{\bm{\pi}}_{i}=\textrm{i}\hbar\nabla_{i}+e\hat{\mathbf{A}}_{\mathrm{ext}}(\hat{\mathbf{r}}_{i})$
are the dynamical momenta, and $\hat{\mathbf{A}}_{\mathrm{ext}}(\hat{\mathbf{r}})=-\bi{e}_{y}B\hat{x}$
describes the magnetic field $\hat{\bi{B}}=\nabla\times\hat{\bi{A}}_{\mathrm{ext}}(\hat{\bi{r}})=B\bi{e}_{z}$.
The cavity field $\hat{\bi{A}}=\sqrt{\hbar/(2\epsilon_{0}\mathcal{V}\omega)}\bi{e}_{y}(\hat{a}+\hat{a}^\dagger)$
is characterized by the in-plane polarization vector $\bi{e}_{x}$
and the photon's bare frequency $\omega$. Further, $W(\hat{\bi{r}}_{i}-\hat{\bi{r}}_{j})=(4\pi\epsilon_{0})^{-1}|\hat{\bi{r}}_{i}-\hat{\bi{r}}_{j}|$
is the Coulomb interaction between the electrons. The transformations of momenta $\{\nabla_{i}\}$ and the two-body interaction are described in App.~\ref{sec:APP-ionsApp}. 

The interaction between the cavity field and the electrons is given by $\hat{\mathbf{A}}\cdot\sum_{i=1}^{N}\textrm{i}\hbar\nabla_{i}+e\hat{\mathbf{A}}_{\mathrm{ext}}(\hat{\mathbf{r}}_{i})=\sqrt{N}\hat{\mathbf{A}}\cdot[\textrm{i}\hbar\nabla_{\mathbf{R}}+e\hat{\mathbf{A}}_{\textrm{ext}}(\hat{\mathbf{R}})]$, such that the cavity field couples only to the CM of the 2deg (App.~\ref{sec:APP-Landau}). The Hamiltonian in the new frame, $\hat{H}_{\textrm{2deg}}=\hat{H}_{\text{2deg-cm}}+\hat{H}_{\text{2deg-rel}}$ is
a sum of (i) the CM part, $\hat{H}_{\textrm{2deg-cm}}$ including the coupling to the quantized field $\hat{\mathbf{A}}$, and (ii) $\hat{H}_{\text{2deg-rel}}$ depending on the relative distances and decoupled from $\hat{\mathbf{A}}$.
The CM part, governing the light-matter dynamics, is given by
\begin{align}
\hat{H}_{\text{2deg-cm}}=\frac{1}{2m}\left(\mathrm{i}\hbar\nabla_{\mathbf{R}}+e\hat{\mathbf{A}}_{\mathrm{ext}}(\hat{\mathbf{R}})+e\sqrt{N}\hat{\mathbf{A}}\right)^{2}+\hbar\omega\left(\hat{a}^\dagger\hat{a}+\frac{1}{2}\right).
\end{align}
The expression for $\hat{H}_{\text{2deg-rel}}$ is presented
in the App.~\ref{sec:APP-Landau}. In the Landau gauge,
the Hamiltonian is translationally invariant along the $y$ axis.
This implies that the eigenfunctions related to this direction in space are plane waves $\exp(\mathrm{i}K_{y}\hat{Y})$.
Applying $\hat{H}_{\text{2deg-cm}}$ to the plane wave and
introducing the shifted coordinate $\hat{\bar{X}}=\hat{X}+\hbar K_{y}/eB$, the Hamiltonian becomes
\begin{equation}
\hat{H}_{\text{2deg-cm}}=-\frac{\hbar^{2}}{2m}\frac{\partial^{2}}{\partial\bar{X}^{2}}+\frac{m\omega_{c}^{2}}{2}\hat{\bar{X}}^{2}-\frac{e^{2}B\sqrt{N}}{m}\hat{A}\hat{\bar{X}}+\frac{e^{2}N\hat{A}^{2}}{2m}+\hbar\omega\left(\hat{a}^\dagger\hat{a}+\frac{1}{2}\right).
\end{equation}
\sloppy To simplify the Hamiltonian further, we perform a Fourier transformation on the electronic coordinate
$\phi(\hat{\bar{X}})=(2\pi)^{-1}\int_{-\infty}^{\infty}\tilde{\phi}(\hat{K})\exp(-\textrm{i}\hat{K}\hat{\bar{X}})\,d\hat{K}$, such that
\begin{equation}
\hat{H}_{\text{2deg-cm}}=-\frac{m\omega_{c}^{2}}{2}\frac{\partial^{2}}{\partial K^2}+\frac{\hbar^{2}}{2m}\hat{K}^{2}+\textrm{i}\frac{e^{2}B\sqrt{N}}{m}\hat{A}\partial_{K}+\frac{e^{2}N\hat{A}^{2}}{2m}+\hbar\omega\left(\hat{a}^\dagger\hat{a}+\frac{1}{2}\right).
\end{equation}
Finally, to turn the Hamiltonian into HH, we introduce annihilation and creation operators $\{\hat{l},\hat{l}^{\dagger}\}$ for the matter degrees of freedom 
$\hat{K}=\sqrt{m \omega_{c}/(2\hbar)}(\hat{l}+\hat{l}^{\dagger})$ and $\partial_{K}=\sqrt{\hbar/(2m\omega_{c})}(\hat{l}-\hat{l}^{\dagger})$, such that
\begin{equation}
\hat{H}_{\textrm{2deg-cm}}=\hbar\omega_{c}\left(\hat{l}^{\dagger}\hat{l}+\frac{1}{2}\right)+\textrm{i}\hbar \sqrt{\frac{ e^2 N \omega_c}{4m\epsilon_{0}\mathcal{V}\omega}}(\hat{a}+\hat{a}^\dagger)(\hat{l}-\hat{l}^\dagger)+\frac{\hbar e^{2}N}{4m\epsilon_{0}\mathcal{V}\omega}\left(\hat{a}+\hat{a}^\dagger\right)^{2}+\hbar\omega\left(\hat{a}^\dagger\hat{a}+\frac{1}{2}\right).
\label{eq:2degHopfield}
\end{equation}
Thus, the CM Hamiltonian for Landau levels coupled to the cavity assumes the same mathematical form as that of harmonically trapped ions, with one important difference.
To obtain Eq.~\eqref{eq:2degHopfield}, we applied a Fourier transformation on the matter operators.
As a result, the dynamical phenomena that occur in real space for the ions manifest in $k$-space for the Landau levels.

\subsection{Harmonic Molecules}

As another example, consider a system of $N$ identical non-interacting polar molecules collectively coupled to a single-mode cavity. The molecular vibrations are described by one-dimensional harmonic potential. For simplicity, the molecules are considered oriented along one of the cavity polarization directions.
For the molecular system, we use the length gauge form of the Pauli-Fierz Hamiltonian~\cite{rokaj2017, ParticlesBook2004}, 
which, given these assumptions, takes the form 
\begin{equation}
\hat{H}_{\mathrm{mol}}=\sum_{i=i}^{N}\left[-\frac{\hbar^{2}}{2M}\frac{\partial^{2}}{\partial x_{i}^{2}}+\frac{M\Omega_{\mathrm{vib}}^{2}\hat{x}_{i}^{2}}{2}\right]+\hbar\omega\left[-\frac{1}{2}\frac{\partial^{2}}{\partial q^{2}}+\frac{1}{2}\left(\hat{q}-g\sum_{i=1}^{N}\hat{x}_{i}\right)^{2}\right].
\end{equation}
Here, $\Omega_{\mathrm{vib}}$ is the fundamental frequency of the harmonic potential, $M$ is the mass of the molecule, and $\omega$
is the cavity frequency. $g=\mu_{0}/\sqrt{\omega\epsilon_{0}\mathcal{V}}$
is the dimensionless light-molecule coupling constant, which depends
on the effective cavity volume $\mathcal{V}$, the vacuum permittivity
$\epsilon_{0}$, and the magnitude of the molecular dipole moment, $\mu_0$.
The operators $\hat{q}$ and $\partial_{q}$ describe the position and momentum quadratures of the bosonic cavity mode. However, they should not be confused with the creation and annihilation photon operators ${\hat{a}^{\dagger}, \hat{a}}$ in the Coulomb gauge, as their connection is subtle~\cite{schaefer2020}.

The model Hamiltonian $\hat{H}_{{\rm \mathrm{mol}}}$ has been used
in multiple works studying molecular systems under vibrational strong
coupling in cavities~\cite{Yang2021, JoelReview, HuoReview2023, horak2024, JoelPolCond}.
The perfectly oriented molecules couple to the cavity through the collective dipole moment. Thus, we transform
$\hat{H}_{{\rm \mathrm{mol}}}$ into the CM frame by introducing the
CM coordinate $\hat{X}=(N)^{-1/2}\sum_{i}\hat{x}_{i}$ and the relative bond lengths $\hat{\tilde{x}}_{j}=(\hat{x}_{1}-\hat{x}_{j})/\sqrt{N}$
with $j>1$. We
already showed how the kinetic energy terms and the harmonic potential transform into the CM frame in App.~\ref{sec:APP-ionsApp}. 
Thus, it is straightforward to obtain
the CM Hamiltonian, while the relative lengths Hamiltonian separates like in the previous models, 
\begin{equation}
\begin{aligned}
\hat{H}_\text{mol-cm} & =-\frac{\hbar^{2}}{2M}\frac{\partial^{2}}{\partial X^{2}}+\frac{M\Omega_{\mathrm{vib}}^{2}}{2}\hat{X}^{2}+\hbar\omega\left[-\frac{1}{2}\frac{\partial^{2}}{\partial q^{2}}+\frac{1}{2}\left(\hat{q}-g\sqrt{N}\hat{X}\right)^{2}\right],\\
\hat{H}_\text{mol-rel} & =\frac{1}{2m}\sum_{j=2}^{N}\left(\frac{\mathrm{i}\hbar}{\sqrt{N}}\partial_{\tilde{x}_{j}}\right)^{2}-\frac{\hbar^{2}}{2mN}\sum_{j,k=2}^{N}\partial_{\tilde{x}_{j}}\cdot\partial_{\tilde{x}_{k}}+\frac{m\Omega^{2}}{2}N\sum_{j=2}^{N}\hat{\tilde{x}}_{j}^{2}-\frac{m\Omega^{2}}{2}\left[\sum_{j=2}^{N}\hat{\tilde{x}}_{j}\right]^{2}.
\end{aligned}
\end{equation}
The cavity mode quadrature, $\hat{q}$, couples only
to the CM coordinate $\hat{X}$ in $\hat{H}_\text{mol-cm}$. We next expand
$(\hat{q}-g\sqrt{N}\hat{X})^{2}$ and apply Fourier transformation to $\hat{q}$, $\phi(\hat{q})=(2\pi)^{-1}\int_{-\infty}^{\infty}\tilde{\phi}(\hat{p})\exp(\mathrm{i}\hat{p}\hat{q})\,d\hat{p}$.
We find that the cavity-molecules CM Hamiltonian has the same mathematical form as the previously discussed two models,
\begin{equation}
\hat{H}_{\text{mol-cm}}=-\frac{\hbar^{2}}{2M}\frac{\partial^{2}}{\partial X^{2}}+\frac{M\Omega_{\mathrm{vib}}^{2}}{2}\hat{X}^{2}+\frac{\hbar\omega g^{2}N}{2}\hat{X}^{2}+\textrm{i}\hbar\omega g\sqrt{N}\hat{X}\frac{\partial}{\partial p}+\hbar\omega\left(-\frac{1}{2}\frac{\partial^{2}}{\partial p^{2}}+\frac{\hat{p}^{2}}{2}\right).
\end{equation}
Finally, we can write the above Hamiltonian in terms of annihilation
and creation operators, $\hat{m}=\sqrt{M\Omega_{\textrm{vib}}/(2\hbar)}[\hat{X}+\hbar/(M\Omega_{\textrm{vib}})\partial_X]$
for matter, and $\hat{c}=(\hat{p}+\partial_p)/\sqrt{2}$ for light,
\begin{equation}
\hat{H}_{\textrm{mol-cm}}=\hbar\Omega_{\textrm{vib}}\left(\hat{m}^{\dagger}\hat{m}+\frac{1}{2}\right)+\textrm{i}\hbar\omega g\sqrt{\frac{\hbar N}{4M\Omega_{\textrm{vib}}}}(\hat{m}^{\dagger}+\hat{m})(\hat{c}-\hat{c}^{\dagger})+\frac{\hbar^{2}\omega g^{2}N}{4M\Omega_{\textrm{vib}}}(\hat{m}^\dagger+\hat{m})^{2}+\hbar\omega\left(\hat{c}^{\dagger}\hat{c}+\frac{1}{2}\right).
\label{eq:molecular-Hamiltonian}
\end{equation}
The molecule-cavity Hamiltonian above has the standard form of the HH~\cite{Hopfield}. However, the bosonic quadratic term shows up for the matter operators
$\hat{m},\,\hat{m}^{\dagger}$. Thus, the dynamics that appear for photons in the ion-cavity system will manifest for the matter in the molecule-cavity system.

\subsection{Hopfield Hamiltonian for the Center of Mass and Polariton Modes}

In the previous subsections, we demonstrated that, across three cavity QED platforms, the collective coupling of an ensemble of particles (ions, 2degs, molecules) is described by the HH in Eq.~(\ref{eq:ionHam}). In all cases, the bosonic/harmonic 
matter part is given by $\sim\hbar\Omega\hat{b}^{\dagger}\hat{b}$,
the cavity mode is described by $\sim\hbar\omega\hat{a}^{\dagger}\hat{a}$,
the bilinear interaction is $\sim\sqrt{N}(\hat{a}+\hat{a}^{\dagger})(\hat{b}-\hat{b}^{\dagger})$. The last term is the quadratic photon self-interaction
in the Coulomb gauge, $\sim N(\hat{a}+\hat{a}^{\dagger})^{2}$, 
or the matter self-interaction, $\sim N(\hat{m}+\hat{m}^{\dagger})^{2}$, in the length gauge.

In what follows, we examine the dynamics of the HH, focusing on trapped ions for concreteness, as this system is more intuitive than 2D electron gas (Landau levels) or molecules. 
We assume the cavity consists 
of two flat parallel mirrors with area $\mathcal{A}$, separated by a distance 
$L$, giving an effective cavity volume $\mathcal{V}=\mathcal{A}L$. 
From this, the 
dimensionless form of Eq.~(\ref{eq:ionHam}) is given by
\begin{align}
\hat{\mathcal{H}}& =\frac{1}{2}(\hat{P}^{2}+\hat{X}^{2})+\frac{\gamma}{2}(\hat{p}^{2}+\hat{q}^{2})-\lambda\hat{q}\hat{P}+\frac{\lambda^{2}}{2}\hat{q}^{2},
\label{eq:Hcm-dimensionless}
\end{align}
where $\hat{X}\equiv(\hat{b}^{\dagger}+\hat{b})/\sqrt{2}$ and $\hat{P}\equiv \mathrm{i}(\hat{b}^{\dagger}-\hat{b})/\sqrt{2}$
are the CM position and momentum operators ($b^{\dagger}$ and $b$
are the corresponding creation and annihilation operators), $\hat{q}\equiv(\hat{a}^{\dagger}+\hat{a})/\sqrt{2}$,
$\hat{p}\equiv \mathrm{i}(\hat{a}^{\dagger}-\hat{a})/\sqrt{2}$ represent the
position and momentum field quadratures. Energy, time, length, and
momentum are measured in units of $\hbar\Omega$, $1/\Omega$, $\sqrt{\hbar/(m\Omega)}$, and
$\sqrt{\hbar m\Omega}$, respectively. The field quadratures associated with 
position and momentum are measured in units of $\sqrt{\hbar/(\epsilon_{0}V\Omega)}$ 
and $\sqrt{\hbar\epsilon_{0}V\Omega}$, respectively. The two dimensionless 
control parameters of $\hat{\mathcal{H}}$ are the frequency ratio between 
the cavity and the CM matter excitations, $\gamma$, and the collective 
coupling constant, $\lambda$, defined as \cite{Vasilis2023}
\begin{equation}
\gamma \equiv \frac{\omega}{\Omega},
\qquad
\lambda \equiv\sqrt{\frac{N g_0^2}{\pi\epsilon_{0}c\mathcal{A}m\Omega}}.
\label{eq:defs-gamma-lamda}
\end{equation}
The two polariton branches have energies $\Omega_{\pm}$,
given by \cite{Vasilis2023}
\begin{equation}
\Omega_\pm^2=\frac{1+\gamma^2+\gamma\lambda^2}{2}
\pm\frac{1}{2}\sqrt{(1+\gamma^2+\gamma\lambda^2)^2-4\gamma^2}.
\label{eq:polariton-freqs}
\end{equation}
The cavity resonates with the CM matter excitations at $\gamma=1$. 
App.\;\ref{sec:APP-Frequency-and-period} shows the 
$\lambda,\gamma$-dependence of $\Omega_\pm$.

We numerically solve the time-dependent Schr\"odinger equation 
with $\hat{\mathcal{H}}$ to study the light-matter dynamics.
We also employ a semi-classical analysis to derive exact
analytical expressions for the observables. This approach 
relies on solving Hamilton's equations of motion, derived 
from the classical equivalent of $\hat{\mathcal{H}}$, for 
$X(t)$, $P(t)$, $q(t)$, and $p(t)$, and provides further 
insight into the underlying dynamics of the coupled system (App.~\ref{sec:APP-Semiclassical-approach}).
According to the Ehrenfest theorem, in a harmonic
system, the first moments $(\text{{e.g.,}}\,\braket{\hat{X}(t)},\,
\braket{\hat{q}(t)},\,\text{etc.})$ match the corresponding 
classical solutions exactly \cite{ShankarBook, TannorBook}. 
Position-momentum uncertainty manifests in higher-order observables 
$(\text{{e.g.,}}\;\braket{\hat{X}^{2}},\;\braket{\hat{P}^{2}},\;
\braket{\hat{q}^{2}},\;\text{{etc.}})$. These observables are obtained 
by averaging classical expressions over the initial phase space 
distribution, corresponding to the composite system initial wave 
function. The Mandel $Q$ functions of matter or light, which quantify 
deviations from classicallity, are obtained using fully quantum simulations.

\section{Scheme I. Transfer of nonclassicality from matter to light}

\begin{figure*}[t]
\begin{centering}
\includegraphics{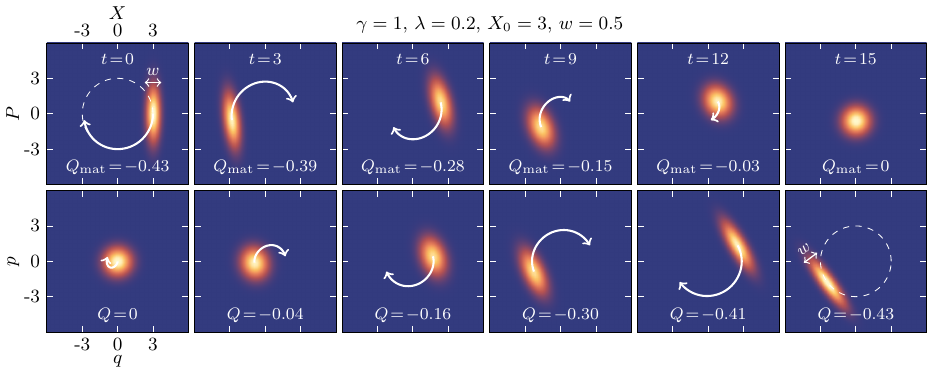} 
\par\end{centering}
\caption{Snapshots of time evolution of the matter ($X,P$, upper row) and cavity 
states ($q,p$, bottom row) in their respective phase spaces. The initial state in Scheme I is 
a product of a displaced quadrature-squeezed matter CM state and cavity ground
state [see Eq.\;\eqref{eq:classical-P}]. Arrows indicate the
trajectories of the phase space distribution centers. 
Initially, the radius of the trajectory equals the initial matter CM quadrature 
shift, $X_{0}$. The dimensional units for the variables and parameters 
in this work are defined after Eq.\;\eqref{eq:Hcm-dimensionless}.}
\label{fig:FIG2} 
\end{figure*}

The transfer of nonclassical states is demonstrated by initializing
the composite system with the matter CM displaced by $X_{0}$ and
quadrature-squeezed in $X$, while the cavity is in the vacuum state.
The initial wave function of the composite system is given by $\psi_{0}(X,q)\propto\exp[-(X-X_{0})^{2}/(2w^{2})]\exp(-q^{2}/2)$, where $w$ is the width of the matter wave function. For $w=1$, 
the matter is in its ground state, and for $w<1$, it is 
quadrature-squeezed. The wave function $\psi_{0}$ corresponds to the 
semi-classical phase space distribution 
\begin{equation}
\mathcal{P}(0)=\frac{1}{\pi^{2}}e^{-[X(0)-X_{0}]^{2}w^{-2}-P^{2}(0)w^{2}}e^{-q^{2}(0)-p^{2}(0)},
\label{eq:classical-P}
\end{equation}
with $X(0)$, $P(0)$, $q(0)$, and $p(0)$ representing the initial conditions.
Fig.\;\ref{fig:FIG2} presents a sequence of phase space snapshots
for the light and matter subsystems. The initial state parameters
are $X_{0}=3$ and $w=0.5$, with light and matter at resonance
($\gamma=1$), in the ultrastrong coupling regime ($\lambda=0.2$).

On short timescales, the squeezed matter phase space distribution
rotates counterclockwise at a radial distance of approximately $X_{0}$
while maintaining its elongated shape. As time progresses, the distribution
becomes circularly symmetric and converges toward the origin. 
The motion of the phase space distribution has two characteristic frequencies defined by the polariton energies in Eq.\;\eqref{eq:polariton-freqs}: fast angular (or rotational) frequency, proportional to $\Omega_+ + \Omega_-$, and slow radial oscillation frequency, proportional to the vacuum Rabi splitting (VRS) $\bar{\Delta}\equiv\Omega_+ - \Omega_-$.

The matter CM dynamics, mediated by the light-matter coupling, generates
photons and the nonclassical state of matter effectively transfers
to the light on the time scale of slow radial oscillation ($\bar{\Delta}^{-1}$), as illustrated in the bottom row in Fig.\;\ref{fig:FIG2}.
The light distribution shifts away from the origin and
becomes squeezed, mirroring the initial state of the matter. 
At resonance, a complete transfer does not require ultrastrong light-matter interaction, but the transfer rate is defined by the coupling strength, $\bar{\Delta}=\lambda$.

More generally, the phase space distributions here are multivariate Gaussians, appearing
as rotated ellipses centered around $(\braket{\hat{X}(t)},\braket{\hat{P}(t)})$
or $(\braket{\hat{q}(t)},\braket{\hat{p}(t)})$. The eigenvectors
of the covariance matrix define the orientation of the ellipses while 
the aspect ratio is proportional to the ratio of the eigenvalues 
(App.\;\ref{sec:APP-Probability-density}).

The Mandel $Q$ function measures the deviation of PND from the Poisson
distribution and is defined as~\cite{KnightBook2012} 
\begin{equation}
Q\equiv\frac{\braket{(\Delta\hat{n})^{2}}-\braket{\hat{n}}}{\braket{\hat{n}}},
\label{eq:Q-def}
\end{equation}
where $\hat{n}=\hat{a}^\dagger\hat{a}=(\hat{q}^2+\hat{p}^2)/2-1/2$,
and $\braket{(\Delta\hat{n})^{2}}\equiv\braket{\hat{n}^{2}}-\braket{\hat{n}}^{2}$. For $Q=0$, photons are in a coherent state and obey trivial Poisson statistics, while for $Q>0$, they obey super-Poissonian statistics, which have a classical analogue. When  $Q<0$, the photons follow sub-Poissonian statistics, a hallmark of quantum light with no classical counterpart~\cite{Mandel1979, DavisDownconversion, KnightBook2012}. 
Under certain conditions--specifically, in a single-mode cavity when the field is in a stationary state, the Mandel $Q$ function is related to the zero-time-delay second-order correlation function $g^{(2)}(0)$ by $
Q = \braket{\hat{n}}[ g^{(2)}(0) - 1]$~\cite{WallsMilburnBook2012, KnightBook2012}. The quantity $g^{(2)}(0)$ is frequently used to infer photon correlations and quantifies the likelihood of simultaneous photon detection events. Here, however, we focus exclusively on the Mandel $Q$ parameter as a measure of the spread of the photon number distribution and do not attempt to relate it to $g^{(2)}(0)$.

Note that quadrature squeezing without displacement
in the phase space does not result in sub-Poissonian statistics \cite{KnightBook2012}.
Geometrically, $\braket{(\Delta\hat{n})^{2}}\leq\braket{\hat{n}}$
occurs when the radial variance of the phase space distribution is
lower than the average radial displacement. In Fig.\;\ref{fig:FIG2},
the \emph{matter} is initially squeezed with $Q_\mathrm{mat}<0$. 
Note that $Q_\mathrm{mat}$ is also defined by Eq.~\eqref{eq:Q-def}, 
but with $\hat{n}=\hat{b}^\dagger\hat{b}=(\hat{X}^2+\hat{P}^2)/2-1/2$ 
being the number operator for the matter CM states. 
While matter and light states exchange, the nonclassical features 
emerge in the light degree of freedom, manifesting in $Q<0$.
The effective transfer of the matter state, characterized by a negative 
$Q$ parameter, into the cavity mode, is one of the key findings of this work.

\subsection{Beating and light-matter resonance}

\begin{figure}
\begin{centering}
\includegraphics{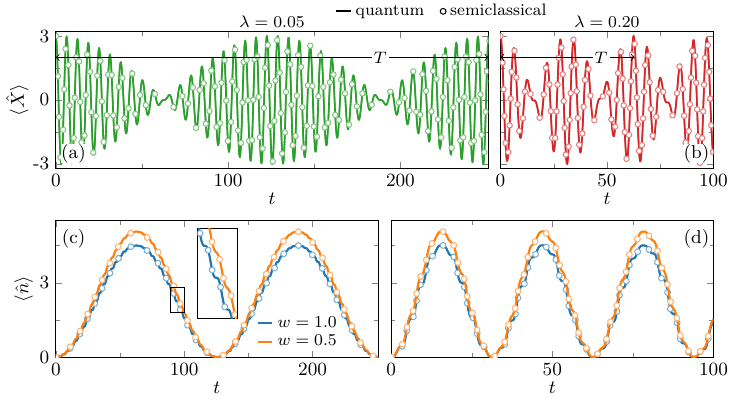} 
\par\end{centering}
\caption{(a,b)\;Expectation values of the matter CM quadrature at resonance, $\braket{\hat{X}(t)}$
[$\text{see\;Eq.\;\eqref{eq:<x(t)>-matter},}\;\gamma=1,\,X_{0}=3,\,w=0.5$]
for two light-matter coupling regimes, strong and ultrastrong. $\braket{\hat{X}(t)}$
is independent of $w$. The beating period is defined as $T\equiv2\pi/(\bar{\Delta}/2)$.
At resonance, $\bar{\Delta}=\lambda$, and $T=4\pi/\lambda\approx251,63$
for (a) and (b), respectively, illustrating the strong dependence
on the collective coupling, $\lambda$. 
(c,d)\;The generated photon number expectation values at resonance
for squeezed and non-squeezed matter. The inset in (c) shows the presence
of high-frequency oscillations. The quantum and semiclassical
results agree in all cases.}
\label{fig:FIG3} 
\end{figure}

Next, we explore the dynamics of the subsystems in more detail. Fig.\;\ref{fig:FIG3}(a,b)
shows the expectation value of
the matter CM position quadrature, $\braket{\hat{X}(t)}$,
at resonance with the cavity ($\gamma=1$), in strong (ultrastrong)
light-matter coupling  regime, $\lambda=0.05$ ($\lambda=0.2$). According
to the Ehrenfest theorem \cite{ShankarBook, TannorBook}, $\braket{\hat{X}(t)}$
equals the classical solution $X(t)$ of a single particle with initial
position $X_0$, and zero momentum (App.\;\ref{sec:APP-Semiclassical-approach}) 
\begin{align}
\frac{\braket{\hat{X}(t)}}{X_{0}}=\cos\left[\frac{\bar{\Sigma}t}{2}\right]\cos\left[\frac{\bar{\Delta}t}{2}\right]+\beta\sin\left[\frac{\bar{\Sigma}t}{2}\right]\sin\left[\frac{\bar{\Delta}t}{2}\right],
\label{eq:<x(t)>-matter}
\end{align}
where $\bar{\Sigma}\equiv\Omega_{+}+\Omega_{-}$, $\bar{\Delta}\equiv\Omega_{+}-\Omega_{-}$
is the vacuum Rabi splitting (VRS), $\bar{\Delta}\leq\bar{\Sigma}$,
and $\beta=(\Omega_{+}^{2}+\Omega_{-}^{2}-2)/(\bar{\Sigma}\bar{\Delta})$.
The higher frequency, $\bar{\Sigma}$, defines the rotation period
of the phase space distributions in Fig.\;\ref{fig:FIG2}. The VRS, on
the other hand, defines the long beating period $T=2\pi/(\bar{\Delta}/2)$
during which $\braket{\hat{X}}$ exhibits slow beatings, and state
transfer occurs. This emergent long timescale, proportional to the VRS, 
depends on the collective coupling and, consequently, the number of particles, $\lambda\propto\sqrt{N}$ (at resonance, $\bar{\Delta}=\lambda$). 

In addition to the collective nature of the beating phenomenon, it is important to highlight the dependence of $T$ on the detuning. In Fig.~\ref{fig:FIG4}(c), we observe that when light and matter excitations are in resonance ($\gamma=1$), the beating period reaches its maximum, and the respective beating frequency is suppressed. This shows that the collective dynamics of an ensemble of particles resonantly coupled to a cavity can be significantly modified even in vacuum, i.e., without an external driving field. Similar modifications of dynamics at resonance have been observed in chemical reactions of collectively coupled molecules under VSC~\cite{Hutchison2012, Hutchison2013, Ebbesen2016, GarciaVidal2021, Sun2024, Xiang2024}. As described in Section~\ref{sec:Hopfield-Hamiltonian}, the HH also applies to harmonic molecules strongly coupled to a cavity mode. Thus, our work could hint at the observed resonant modification of reactions in polaritonic chemistry~\cite{HuoReview2023, JoelReview, GarciaVidal2021}. Notably, the beating period exhibits a sharp resonance in the strong coupling regime ($\lambda=0.05$), while in the ultrastrong regime ($\lambda=0.2$), the peak broadens
as shown in Fig.~\ref{fig:FIG4}(d). This is due to CRT, on which we focus next.

\subsection{Effects of the Counter-rotating Terms on Dynamical Observables}

\begin{figure}
\begin{centering}
\includegraphics{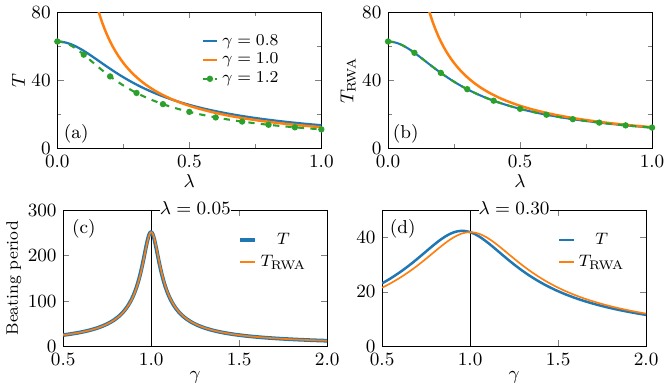} 
\par\end{centering}
\caption{Dependence of the beating periods, $T\equiv4\pi/(\Omega_{+}-\Omega_{-})$
[Eq.\;\eqref{eq:polariton-freqs}] and $T_{\mathrm{RWA}}\equiv4\pi/(\Omega_{\mathrm{RWA,}+}-\Omega_{\mathrm{RWA},-})$
[Eq.\;\eqref{eq:APP-polariton-branches-RWA}] on (a,b) $\lambda$
and (c,d) $\gamma$. $T$ has a distinctive asymmetry as
the cavity frequency is scanned through resonance. In contrast, $T_\text{RWA}$ is independent of the sign of the detuning,
such that the $\gamma=0.8,1.2$ curves in (b) overlap, and the RWA
curve in (d) is symmetric about the point $\gamma=1$.}
\label{fig:FIG4} 
\end{figure}

The CRT make the polariton avoided crossing sensitive to the sign of the relative detuning, $\gamma-1=(\omega-\Omega)/\Omega$,
and the difference between red- and blue-shifted cavity cases increases
with $\lambda$ (App. \ref{sec:APP-Frequency-and-period}). This is a characteristic of the ultrastrong coupling regime~\cite{kockum2019ultrastrong, UltrastrongReview2019}. Consequently, the beating period, $T$, is also sensitive
to the detuning sign, as shown in Fig.\;\ref{fig:FIG4}(a). In contrast, under
the rotating wave approximation (RWA), when the terms proportional
to $\lambda\hat{a}^{\dagger}\hat{b}^{\dagger},\,\lambda\hat{a}\hat{b}\;\text{and}\;\lambda^{2}$
are omitted from Eq.\;\eqref{eq:Hcm-dimensionless}], $\bar{\Delta}$
is independent of the detuning sign. 
Fig.\;\ref{fig:FIG4}(b) shows $T_{\mathrm{RWA}}$, which remains the same 
for detunings of either sign. 

Fig.\;\ref{fig:FIG4}(c) demonstrates that the RWA 
is a good approximation for $\lambda<0.1$, as $T\approx T_{\mathrm{RWA}}$, 
which is expected. 
However, the asymmetry of $T$ around $\gamma = 1$ becomes evident in panel (d).
Additionally, the prefactor $\beta$ in Eq.\;\eqref{eq:<x(t)>-matter}
vanishes at resonance under the RWA (App.\;\ref{sec:APP-Semiclassical-approach}).
The asymmetries in the time-dependent matter CM quadrature are dynamic
manifestations of the polariton branch asymmetry around $\gamma = 1$.
Although the asymmetries of $\braket{\hat{X}(t)}$ 
arise from ultrastrong coupling, they are not inherently quantum effects. 
As mentioned earlier, in a system of coupled harmonic oscillators, the 
normal-mode frequencies, $\Omega_\pm$, and the first moments are 
determined by the classical solution, meaning similar features can be 
observed even in a pair of coupled classical oscillators.
Therefore, next, we consider time-dependent observables defined in
terms of the second moments of the light probability distribution 
to identify potential quantum effects.

\begin{figure}
\begin{centering}
\includegraphics{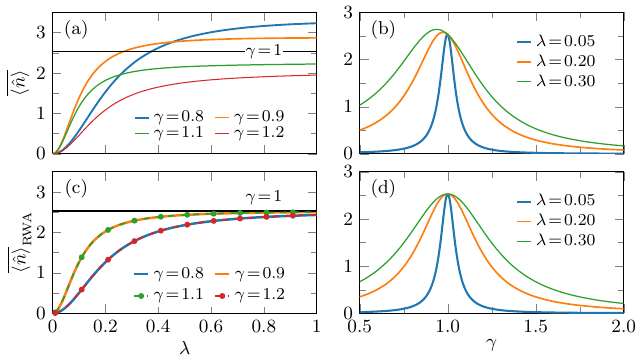} 
\par\end{centering}
\caption{(a,b) Dependence of average photon number in
Eq.\;\eqref{eq:n-time-averaged} on $\lambda$ and $\gamma$. (c,d) The same as in (a,b), but under RWA.
At resonance, the average number of photons becomes independent of the coupling
strength, $\lambda$ in both the full system and under RWA--all graphs in panels (b)
and (d) overlap at $\gamma=1$.
Note that in panel (d), the functions are symmetric Lorentzians centered
at $\gamma=1$.}
\label{fig:FIG5} 
\end{figure}

Fig.\;\ref{fig:FIG3}(c,d) shows the evolution of the photon number
expectation value, $\braket{\hat{n}(t)}$ at resonance ($\gamma=1$).
The local maxima correlate to when the phase space distribution is
farthest from the origin. The time-averaged expectation value 
is given by (for the full expression, see App.\;\ref{sec:APP-Photon-number-expectation})
\begin{equation}
\overline{\braket{\hat{n}}}
\approx
\Delta\overline{\braket{\hat{n}}}
=
\lambda^2 f_1 (w^{-2} - 1) + \lambda^2 f_2 (2X_0^2 + w^2 - 1),
\label{eq:n-time-averaged}
\end{equation}
where $\Delta\braket{\hat{n}(t)}\equiv\braket{\hat{n}(t)}-\braket{\hat{n}(t;X_{0}=0,w=1)}$.
To simplify the expression, we subtract the relatively small contribution from the average photon number when the CM
is in the ground state. Here, $f_{1,2}$
are functions of $\Omega_{\pm}$. At resonance, $\Delta\overline{\braket{\hat{n}(\gamma=1)}}=X_{0}^{2}/4+(w-w^{-1})^{2}/8$.
For negligible matter squeezing ($w\approx1$), $\Delta\overline{\braket{\hat{n}}}\approx(2\lambda^{2}f_{2})X_{0}^{2}$,
and the average number of generated photons is determined by the initial
CM shift, $X_{0}$. Fig.\;\ref{fig:FIG5}(a,b) shows the dependence
of $\Delta\overline{\braket{\hat{n}}}$ on $\lambda$ and $\gamma$.
The average photon number is asymmetric relative to the resonance point and
fewer photons are generated, on average, for $\gamma>1$. This asymmetry
has important implications for the photon number distribution as 
well because, as we discuss next, the $Q$ function is more negative in 
the blue-shifted cavity.

Under RWA, $\overline{\braket{\hat{n}}}_{\mathrm{RWA}}=\lambda^{2}f_{\mathrm{RWA}}[2X_{0}^{2}+(w-w^{-1})^{2}]$,
and $f_{\mathrm{RWA}}=1/[8(\gamma-1)^{2}+8\lambda^{2}]$ is symmetric
about $\gamma=1$ (App.\;\ref{sec:APP-Photon-number-expectation}).
Fig\;\ref{fig:FIG5}(c,d) shows the dependencies of $\overline{\braket{\hat{n}}}_{\mathrm{RWA}}$.
In panel\;(c), the $\gamma=0.8,1.2$ and $\gamma=0.9,1.1$ curves
overlap, while panel\;(d) demonstrates the symmetric Lorentzian function.
The blue curves ($\lambda=0.05$) in panels (b) and (d) are practically
indistinguishable, suggesting that the RWA is a good approximation
in the weak coupling regime ($\lambda\ll1$). 

The asymmetries in the time-averaged photon number reflect the effects of ultrastrong light-matter coupling. In contrast to the first 
moments, the photon number explicitly depends on the quantum uncertainty in
position- and momentum-related quadratures. The average photon number is 
an experimentally accessible observable (for details, see Sec.\;\ref{sec:Experimental-observation})
and thus allows probing effects beyond RWA in quantum systems.

\subsection{Mandel \protect{$Q$} Function}

\begin{figure}
\begin{centering}
\includegraphics{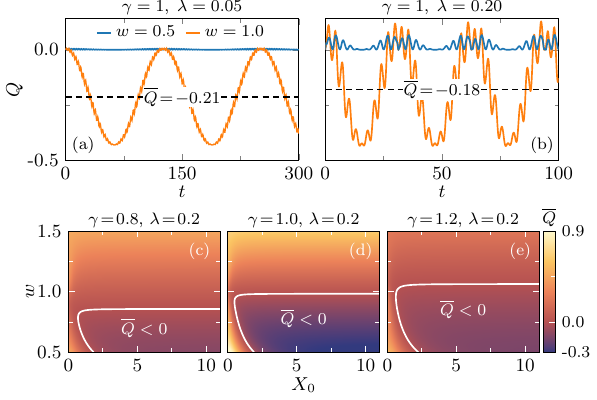} 
\par\end{centering}
\caption{(a,b)\;Mandel $Q$ functions for
the same $w$ and $\lambda$ values. Importantly, for $w=0.5$, the
time-averaged $Q$ function, $\overline{Q}$ attains negative values.
(c-e) The dependence of photon $\overline{Q}$ on the initial
displacement and width of the matter CM state is shown before (c), at (d),
and after (e) the cavity resonance for strong cavity coupling. The nonclassical features of the photon state primarily occur for $w<1$. At resonance
in (d), the state transfer is most efficient. The asymmetry between
(c) and (e) vanishes in the RWA (see App.\;\ref{sec:APP-Mandel-Q-function}),
and $\overline{Q}<0$ can be achieved only for $w<1$.}
\label{fig:FIG6} 
\end{figure}

Next, we focus on the Mandel $Q$ function, which quantifies nonclassicality, indicated by negative values of $Q$. Fig.\;\ref{fig:FIG6}(a,b) illustrates the evolution 
of the photon $Q$ function at resonance, when $Q$ attains negative values only when
CM is both shifted \emph{and} quadrature-squeezed in $X$ (i.e., when $w<1$). Qualitatively,
Fig.\;\ref{fig:FIG2} suggests that the minimum of $Q$ is connected
to the initial value of $Q_{\mathrm{mat}}$, given by $Q_{\mathrm{mat}}(0)=(w^{2}-1)[1+\mathcal{O}(X_{0}^{-2})]$
(App.\;\ref{sec:APP-Mandel-Q-function}). While $\overline{\braket{\hat{n}}}$
in Eq.\;\eqref{eq:n-time-averaged} scales with $X_{0}^{2}$, the
initial $Q_{\mathrm{mat}}$ approaches a constant as $X_{0}$ increases,
leading to the saturation of the minimal attainable $Q$ values.

The time-averaged photon Mandel function, $\overline{Q}\equiv\lim_{\tau\rightarrow\infty}\tau^{-1}\int_{0}^{\tau}Q(t)\,dt$,
is shown in Fig.\;\ref{fig:FIG6}(c-e) for a range of $X_{0}$ and
$w$ values. $\overline{Q}$ is negative on either side of the resonance for 
an initially shifted and quadrature-squeezed matter state.
Notably, $Q$ may attain negative values for an initially quadrature-squeezed
matter state even without the diamagnetic term ($\propto\lambda^{2}$)
and under the RWA. However, under the RWA, the asymmetry between
(c) and (e) disappears, and negative $\overline{Q}$ can be achieved
only for $w<1$ (App.\;\ref{sec:APP-Mandel-Q-function}).

\section{Scheme II. generation of  nonclassicality from a coherent cavity}

\begin{figure*}[t]
\begin{centering}
\includegraphics{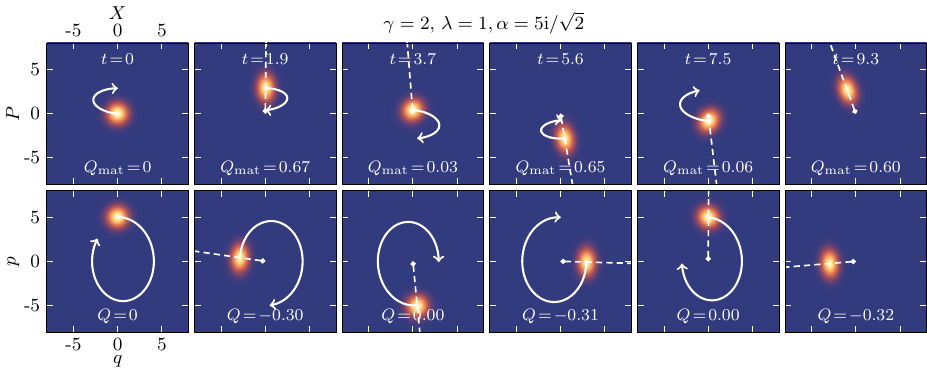} 
\par\end{centering}
\caption{Snapshots of evolving phase space distribution of matter ($X,P$,
upper row) and cavity field ($q,p$, bottom row). The initial state in Scheme II
is a product of matter CM ground state and cavity coherent state, $\ket{\alpha}$.
The arrows show the trajectories of the distribution centers. The
dashed lines pass through the origin and distribution centers in each
panel. Note that the matter distributions approximately align with
the dashed lines, while the light distributions are perpendicular. The values of detuning and coupling constant are
exaggerated to highlight the effect.}
\label{fig:FIG7} 
\end{figure*}

An alternative approach for generating nonclassical states involves initializing
the system with the matter CM in its ground state and the cavity in
a coherent state, $\ket{\alpha}$. Fig.\;\ref{fig:FIG7} illustrates
the evolution of the phase space distributions of matter CM (top row)
and light (bottom row) in Scheme II. Initially, both $Q$ and $Q_{\mathrm{mat}}$
are zero, as the system begins in a trivial state without any 
correlations. In this scenario, the counter-rotating interaction terms are 
crucial in generating sub-Poissonian distributions in the motional states
of matter or photon number states (App.\;\ref{sec:APP-Mandel-Q-function}).

Over time, the matter and light distributions periodically elongate 
(squeeze); however, during maximal elongation, the matter phase space 
distribution aligns approximately along the line connecting its center 
to the origin, while the light distribution is oriented perpendicularly.
As previously noted, the
$Q$ function becomes negative when the radial variance of the phase
space distribution is less than the radial displacement. In this scenario, only the photon $Q$ function attains negative values. In contrast to the resonance case shown in Fig.\;\ref{fig:FIG2},
the state transfer is only partial due to the large detuning. The
matter phase space distribution doesn't reach the initial radial displacement
of the light distribution, and the light distribution does not collapse
to the origin.

\begin{figure}
\begin{centering}
\includegraphics{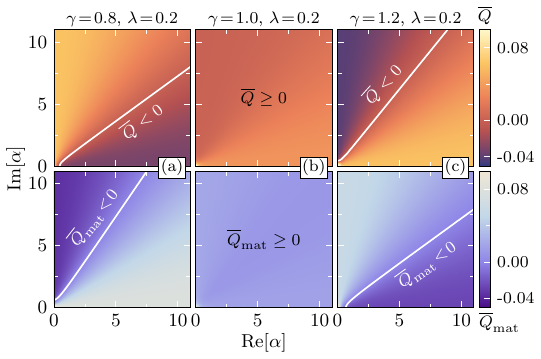} 
\par\end{centering}
\caption{The dependence of photon $\overline{Q}$ (top panels)
and the matter $\overline{Q}_{\mathrm{mat}}$ (bottom
panels) on the initial coherent state, $\ket{\alpha}$. Sub-Poissonian
distributions of photon and motional
states are achieved off-resonance in (a) and (c). With increasing
$\mathrm{Re}[\alpha]$
 and $\mathrm{Im}[\alpha]$ components, the sub-Poissonian distributions saturate. At resonance, both photon and matter Mandel parameters are positive.}
\label{fig:FIG8} 
\end{figure}

Fig.\;\ref{fig:FIG8} shows the time averaged  photon
$\overline{Q}$ (top panels) and matter $\overline{Q}_{\mathrm{mat}}$
(bottom panels). $\overline{Q}$ is sensitive to
the phase of $\alpha$, negative $\overline{Q}$ is obtained before 
(after) the resonance for coherent states with sufficiently large $\mathrm{Re}[\alpha]$
($\mathrm{Im}[\alpha]$). The $Q$ functions reach saturation along
the real or imaginary axes. At resonance, both $\overline{Q}$ and $\overline{Q}_{\mathrm{mat}}$ remain non-negative.
Additionally, a clear asymmetry appears between the red and blue detuned cavities in panels (a) and (c).
This asymmetry increases with $\lambda$, and, in the case of photons, culminates with the complete absence of $\overline{Q}<0$
in the red-shifted cavity in the ultra-strong coupling regime ($\lambda=0.5$,
see App.\;\ref{sec:APP-Mandel-Q-function}).

Remarkably, under initial conditions where the photon $\overline{Q}\geq0$
in Fig.\;\ref{fig:FIG8}(a,c), $\overline{Q}_{\mathrm{mat}}$ attains
negative values. In other words, even when the sub-Poissonian PND
is not achieved, the coherent state is partially transferred to the
matter subsystem, simultaneously developing the sub-Poissonian distribution
of motional states of the \emph{matter} CM. The generation of nonclassical states in light and matter, starting from a coherent state of light, represents a key finding of this work.

\begin{figure}[h]
\begin{centering}
\includegraphics{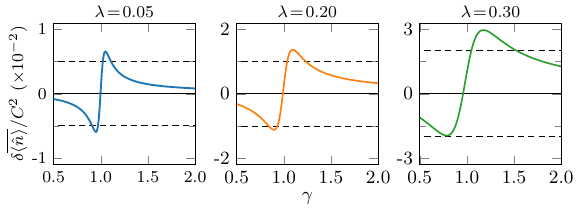} 
\par\end{centering}
\caption{Dependence of $\delta\overline{\braket{\hat{n}}}/C^{2}$ in Eq.\;\eqref{eq:APP-delta-n-Scheme-II}
on $\gamma$. In all cases, on average, the cavity is populated with
more photons starting from a coherent state with imaginary (real)
$\alpha$ below (above) resonance. The asymmetry of $|\delta\overline{\braket{\hat{n}}}|$
about $\gamma=1$ increases with $\lambda$.}
\label{fig:FIG9} 
\end{figure}

The average number of photons, $\overline{\braket{n}}$,
is also sensitive to the phase of $\alpha$ and shows trends similar
to those of $\overline{Q}$. The symmetries of $\overline{\braket{n}}$
can be accessed via $\delta\overline{\braket{\hat{n}}}\equiv\overline{\braket{\hat{n}(\alpha=C)}}-\overline{\braket{\hat{n}(\alpha=\mathrm{i}C)}}$,
where $C>0$. $\delta\overline{\braket{\hat{n}}}$ measures the difference
in average photon number between the initial coherent state with $\mathrm{Re}[\alpha]$ vs.
$\mathrm{Im}[\alpha]$, and it is given by 
\begin{equation}
\delta\overline{\braket{\hat{n}}}=\lambda^{2}C^{2}\frac{(2\gamma+\lambda^{2})(\Omega_{+}^{2}+\Omega_{-}^{2})-4\gamma}{4\bar{\Sigma}^{2}\bar{\Delta}^{2}}.
\label{eq:APP-delta-n-Scheme-II}
\end{equation}
Fig.\;\ref{fig:FIG9} shows the dependence of $\delta\overline{\braket{\hat{n}}}/C^{2}$
on the system parameters. Here, the value of $C$ is large enough
to saturate the $Q$ parameter. On average, the cavity is populated
with fewer photons for $\mathrm{Re}[\alpha]$
($\mathrm{Im}[\alpha]$), before (after) the resonance. 

Moreover, the asymmetry
of $|\delta\overline{\braket{\hat{n}}}|$ increases with $\lambda$. In the rotating wave approximation, $\delta\overline{\braket{\hat{n}}}=0$. To our knowledge, this asymmetry in photon generation as light-matter coupling enters the ultrastrong regime has not been previously discussed. This phenomenon could potentially be tested in Landau polariton systems, where ultrastrong coupling has been demonstrated~\cite{UltrastrongReview2019, kockum2019ultrastrong, Scalari2012, Li2018}. Such observation would provide clear evidence of the dynamical manifestation of ultrastrong coupling, extending our understanding of this regime beyond conventional spectroscopic methods.

\section{How to access the nonclassical states\label{sec:Experimental-observation}}

In Scheme I, the nonclassical photons generated in the cavity can be extracted by making one of the cavity mirrors semi-transparent \cite{Kimble1991, Kimble1997}.
The output field is obtained by employing  the input-output formalism,
which assumes a bilinear coupling to the bath degrees of freedom (cavity
environment) through the field operators $\hat{a}$ and $\hat{a}^{\dagger}$,
with a constant coupling proportional to $\sqrt{\kappa}$. In this framework, the
relationship between the cavity field operator $\hat{a}(t)$ and
the output field operator, $\hat{d}_{\mathrm{out}}(t)$ is given by
\cite{Gardiner1985} 
\begin{equation}
\dot{\hat{a}}=-\mathrm{i}[\hat{a},\hat{\mathcal{H}}]+\frac{\kappa}{2}\hat{a}-\sqrt{\kappa}\hat{d}_{\mathrm{out}}.
\label{eq:da/dt-b_out}
\end{equation}
The bath can also feed energy into the cavity through the operator
$\hat{d}_{\mathrm{in}}(t)$. The input-output relation states that
$\hat{d}_{\mathrm{out}}(t)-\hat{d}_{\mathrm{in}}(t)=\sqrt{\kappa}\hat{a}(t)$.
In practice, the effect of coupling on the system can be formulated
in terms of the master equation for the reduced density matrix of
the system, $\hat{\rho}$. In the low-temperature limit ($k_{B}T\ll\hbar\omega_{i}$,
where $\omega_{i}$ is an eigenfrequency of the system) and weak system-bath
coupling, the master equation becomes \cite{Gardiner1985} 
\begin{equation}
\dot{\hat{\rho}}=-\mathrm{i}[\hat{\mathcal{H}},\hat{\rho}]+\frac{\kappa}{2}(2\hat{a}\hat{\rho}\hat{a}^{\dagger}-\hat{a}^{\dagger}\hat{a}\hat{\rho}-\hat{\rho}\hat{a}^{\dagger}\hat{a}).
\label{eq:master-equation}
\end{equation}

Eq.\;\eqref{eq:master-equation} can be used to write the master
equation for the annihilation operator, $\hat{a}$, $\dot{\hat{a}}=-\mathrm{i}[\hat{a},\hat{\mathcal{H}}]-(\kappa/2)\hat{a}$.
Substituting this into Eq.\;\eqref{eq:da/dt-b_out} yields $\hat{d}_{\mathrm{out}}=\sqrt{\kappa}\hat{a}$. 
Therefore, the number of photons leaking from the cavity is proportional to the number inside, $\braket{\hat{n}_{\mathrm{out}}(t)}=\kappa\braket{\hat{n}(t)}$. 
Similarly, $Q_{\mathrm{out}}(t)=\kappa Q(t)$.

\begin{figure}[h]
\begin{centering}
\includegraphics{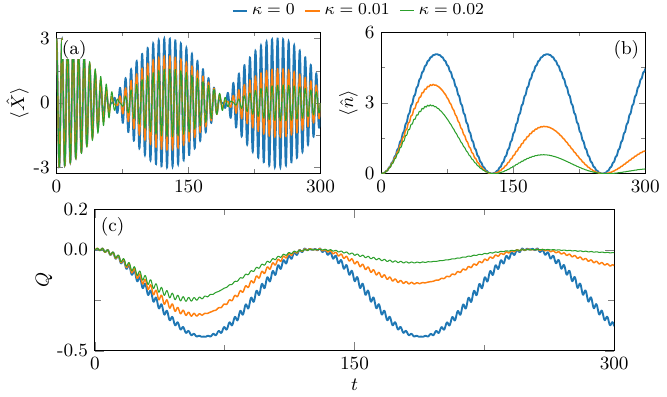} 
\par\end{centering}
\caption{Expectation values of (a)\;matter CM $X$ quadrature, (b)\;photon number,
and (c)\;photon Mandel $Q$ in the presence of cavity loss. 
The results were obtained by numerically solving Eq.\;\eqref{eq:master-equation}.
Parameters: $\gamma=1,\,X_{0}=3,\,w=0.5$.}
\label{fig:FIG10} 
\end{figure}

Fig.\;\ref{fig:FIG10} shows expectation values obtained by numerically
solving Eq.\;\eqref{eq:master-equation}. The observables exhibit persistent oscillations, exponentially decaying at an approximate
rate of $\kappa/2$ (assuming $\kappa\ll\Omega$). The time-averaged expectation value of the photons outside the cavity can be
approximated as $\overline{\braket{\hat{n}_{\mathrm{out}}(t)}}\approx\kappa\overline{\braket{\hat{n}}}\exp(-\kappa t/2)$
where $\overline{\braket{\hat{n}}}$ is given by Eq.\;\eqref{eq:n-time-averaged}.
Constrained optimization can be employed to simultaneously minimize $Q_{\mathrm{out}}$ 
and maximize $\braket{\hat{n}_{\mathrm{out}}}$, thereby optimizing detection efficiency, 
but this is beyond the scope of the present study.

Alternatively, in Scheme II, the sub-Poissonian distribution of the matter CM states can 
be directly observed via spectroscopic methods, such as sideband spectroscopy for trapped 
atomic ions \cite{Wineland96, Hite2013}. Light and matter can be decoupled by introducing 
a large detuning to measure the properties of matter independently. The method and the 
details for introducing detuning will depend on the system and experimental setup.

\section{Conclusions and outlook}

By leveraging the formation of polaritons in strongly and ultrastrongly coupled light-matter systems, we propose two initialization schemes to generate and effectively exchange nonclassical states of matter and light. In the first scheme, an initially displaced and quadrature-squeezed matter CM state is dynamically transferred to the cavity mode, producing nonclassical photons that can be extracted by making one of the cavity mirrors semi-transparent. This scheme builds upon the ability to create coherent and quadrature-squeezed matter states in ion traps (e.g., see~\cite{Wineland96}).

The second initialization scheme relies on the ultrastrong coupling effects (i.e., on the presence of CRT in the Hamiltonian), which generate sub-Poissonian distributions of photon or matter states, starting from a coherent cavity state. This shows that an ultrastrongly coupled light-matter system acts as a generator that transforms a coherent photonic state, devoid of nonclassical characteristics, into a nonclassical state of light or matter. 
This demonstrates a clear quantum advantage of the ultrastrong coupling regime~\cite{kockum2019ultrastrong, UltrastrongReview2019} and suggests novel pathways for hybrid, polaritonic quantum technologies. 
For example, one could envision protocols where matter is prepared in such a state that the system evolves into a macroscopic cat state~\cite{Gao2010} or a photonic cluster state~\cite{photoncluster2024}, which can be used for quantum information and computation. 

On the matter side, the population distribution of the matter states can be measured spectroscopically after decoupling the matter from the cavity. Additional manifestations of the ultrastrong coupling effects are the distinctive asymmetries in dynamical observables, such as average photon number and Mandel $Q$ parameter, emerging as the cavity frequency is scanned through the resonance. These CRT-induced asymmetries provide a novel, dynamical perspective on the ultrastrong coupling regime typically probed by transmission spectroscopy. To our knowledge, such dynamic asymmetries have not been reported elsewhere.

The considered schemes can act as probes of quantum dynamics under strong or ultrastrong light-matter coupling in microscopic and mesoscopic systems. The coupling to light in such systems differs from that in typical optomechanical systems~\cite{AspelmeyerReview2014}. Initializing the cavity field in non-trivial states may enable new quantum control schemes for chemical processes in molecular systems under VSC. This could lead to applications in catalysis or the development of novel reaction pathways influenced by quantum states of light. Similarly, for Landau levels coupled to the cavity mode, strong light-matter interactions can transfer states of light to the electronic subsystem's collective (CM) mode, offering new ways to control electronic properties. Once a strong-ultrastrong coupling regime between the trapped ion’s motion and the cavity mode is achieved, nonclassical photons could be potentially generated from squeezed motional states of ions. Nonclassical photonic states are important for sensing beyond the standard quantum limit~\cite{QSensing, QJumps, Catsensor, XieQuantumLimit}.

We anticipate that our analysis will inspire further experimental and theoretical investigations of the dynamics in various strongly and ultrastrongly coupled light-matter systems, ultimately advancing the field of quantum science and technology and bridging it with polaritonic chemistry~\cite{Ebbesen2016, HuoReview2023, JoelReview} and cavity quantum materials~\cite{Schlawin2022, GarciaVidal2021}.

\begin{acknowledgments}
I.T., V.R., and H.R.S. acknowledge support from the NSF through a
grant for ITAMP at Harvard University. I.T. was also supported by
an NSF subcontract No. 3357 at the Smithsonian Astrophysical Observatory.
J.C. acknowledges support from the NSF (Grants No.\;CHE1800301 and
No.\;CHE2324300), and the MIT Sloan Fund. 
\end{acknowledgments}

\appendix

\section{Mapping to the Hopfield Hamiltonian for Different Systems \label{sec:APP-Harmonic-Cavity-QED}}

In this Appendix, we detail the intermediate steps involved in 
transforming the Hamiltonians of the various systems in Section~\ref{sec:Hopfield-Hamiltonian} into the HH.

\subsection{Cold Trapped Ions \label{sec:APP-ionsApp}}

The collective coupling between the trapped ions and the cavity field emerges by transforming the Hamiltonian in Eq.\;\eqref{eq:Pauli-Fierz} to the CM frame. A similar transformation will also be used for the other systems considered later; so, we will show all the steps here.
For mathematical convenience, we utilize a symmetric definition with respect to $\sqrt{N}$ for the coordinates in the CM frame \cite{BuschSol} 
\begin{equation}
\hat{\mathbf{R}}=\frac{1}{\sqrt{N}}\sum_{i=1}^{N}\hat{\mathbf{r}}_{i}\quad\textrm{and}\quad\hat{\tilde{\mathbf{r}}}_{j}=\frac{\hat{\mathbf{r}}_{1}-\hat{\mathbf{r}}_{j}}{\sqrt{N}}\quad\textrm{with}\quad j>1.
\end{equation}
The momenta of the particles in the new coordinate system are $\nabla_{1}=(\nabla_{\mathbf{R}}+\sum_{j=2}^{N}\widetilde{\nabla}_{j})/\sqrt{N}$
and $\nabla_{j}=(\nabla_{\mathbf{R}}-\widetilde{\nabla}_{j})/\sqrt{N}\;\textrm{with}\;j>1.$
The kinetic energy terms in the new frame are given by
\begin{align}
\sum_{i=1}^{N}\nabla_{i}^{2} & =\nabla_{\mathbf{R}}^{2}+\frac{1}{N}\sum_{j=2}^{N}\widetilde{\nabla}_{j}^{2}+\frac{1}{N}\sum_{j,k=2}^{N}\widetilde{\nabla}_{j}\cdot\widetilde{\nabla}_{k}.
\label{eq:APP-momenta}
\end{align}
In the new coordinates, the cavity field couples only to the CM momentum,
$\hat{\mathbf{A}}\cdot\sum_{i=1}^{N}\nabla_{i}=\sqrt{N}\hat{\mathbf{A}}\cdot\mathbf{\nabla}_{\mathbf{R}}$,
while the scalar trapping potential separates into two parts, one depending on the CM coordinate and the other one on the
relative coordinates, without cross-terms between the two
\begin{equation}
\sum_{i=1}^{N}\hat{\mathbf{r}}_{i}^{2}=\hat{\mathbf{R}}^{2}+N\sum_{j=2}^{N}\hat{\mathbf{R}}_{j}^{2}-\left[\sum_{j=2}^{N}\hat{\mathbf{R}}_{j}\right]^{2}.
\label{eq:APP-potential}
\end{equation}
The two-body interaction $W(\hat{\mathbf{r}}_{i}-\hat{\mathbf{r}}_{l})$
depends only on the relative distances and thereby does not affect
the cavity-induced CM motion. In the CM frame, it is given by
\begin{equation}
\sum\limits _{i<l}^{N}W(\hat{\mathbf{r}}_{i}-\hat{\mathbf{r}}_{l})=\sum_{1<l}^{N}W(\sqrt{N}\hat{\tilde{\mathbf{r}}}_{l})+\sum_{2\leq i<l}^{N}W\left(\sqrt{N}(\hat{\tilde{\mathbf{r}}}_{i}-\hat{\tilde{\mathbf{r}}}_{l})\right).
\end{equation}
The analysis above shows that the Hamiltonian separates into two
parts, $\hat{H}_{\mathrm{ion}}=\hat{H}_{\text{ion-cm}}+\hat{H}_{\text{ion-rel}}$, where (i) $\hat{H}_{\text{ion-cm}}$, describes the CM coupling to the quantized field $\hat{\mathbf{A}}$, and (ii) $\hat{H}_{\text{ion-rel}}$, describes the dynamics of the relative coordinates, decoupled from $\hat{\mathbf{A}}$ and the CM. The two parts are given by 
\begin{equation}
\begin{aligned}
\hat{H}_{\textrm{ion-cm}} & =\frac{1}{2m}(\textrm{i}\hbar\nabla_{\mathbf{R}}+g_{0}\sqrt{N}\hat{\mathbf{A}})^2+\frac{m\Omega^2}{2}\hat{\mathbf{R}}^{2}+\sum_{\nu=x,y}\hbar\omega\left(\hat{a}_\nu^\dagger\hat{a}_\nu+\frac{1}{2}\right),\\
\hat{H}_{\textrm{ion-rel}}= & -\frac{\hbar^{2}}{2mN}\sum_{j=2}^{N}\widetilde{\nabla}_{j}^{2}-\frac{\hbar^2}{2mN}\sum_{j,k=2}^{N}\widetilde{\nabla}_{j}\cdot\widetilde{\nabla}_{k}+\frac{mN\Omega^{2}}{2}\sum_{j=2}^{N}\hat{\tilde{\mathbf{r}}}_{j}^{2} \\
- & \frac{m\Omega^{2}}{2}\left[\sum_{j=2}^{N}\hat{\tilde{\mathbf{r}}}_{j}\right]^{2}+\sum_{1<l}^{N}W(\sqrt{N}\hat{\tilde{\mathbf{r}}}_{l})+\sum_{2\leq i<l}^{N}W\left(\sqrt{N}(\hat{\tilde{\mathbf{r}}}_{i}-\hat{\tilde{\mathbf{r}}}_{l})\right).
\end{aligned}
\end{equation}
In addition, it is crucial to demonstrate that the CM and relative
distance degrees of freedom are independent by checking the commutation
relations between their coordinates and momenta. Using the chain rule,
we find the derivatives in the CM frame 
\begin{equation}
\nabla_{\mathbf{R}}=\frac{1}{\sqrt{N}}\sum_{i=1}^{N}\nabla_{i}\quad\textrm{and}\quad\widetilde{\nabla}_{j}=\frac{1}{\sqrt{N}}\sum_{i=1}^{N}\nabla_{i}-\sqrt{N}\nabla_{j}\quad\text{with}\quad j>1,
\end{equation}
It is clear that the momenta
in the new frame of reference commute, $[\nabla_{\mathbf{R}},\widetilde{\nabla}_{j}]=0$.
It can also be shown that the momenta and coordinates are independent since $[\nabla_{\mathbf{R}},\hat{\tilde{\mathbf{r}}}_{j}]=0$
and $[\widetilde{\nabla}_{j},\hat{\mathbf{R}}]=0$. Thus, we focus on the CM part to describe the cavity-matter dynamics. Since
the polarization vectors of the photon field lie in the $(x,y)$ plane,
the $z$ direction of the system becomes trivial.
The light-matter Hamiltonian then describes a system of interacting
harmonic oscillators, and importantly, the $x$ and
$y$ directions are independent, 
\begin{equation}
\hat{H}_{\text{ion-cm}}=\sum_{\nu=x,y}\Big[-\frac{\hbar^{2}}{2m}\partial_{R_{\nu}}^{2}+\frac{\textrm{i}g_{0}\hbar}{m}\sqrt{N}\hat{A}_{\nu}\partial_{R_{\nu}}+\frac{m\Omega^{2}}{2}\hat{R}_{\nu}^{2}+\frac{Ng_{0}^{2}}{2m}\hat{A}_{\nu}^{2}+\hbar\omega\left(\hat{a}_{\nu}^{\dagger}\hat{a}_{\nu}+\frac{1}{2}\right)\Big].
\end{equation}
To avoid any confusion, note that $\hat{\mathbf{R}}=(\hat{R}_{x},\hat{R}_{y})=(\hat{X},\hat{Y})$,
$\nabla_{\mathbf{R}}=(\partial_{R_{x}},\partial_{R_{y}})=(\partial_{X},\partial_{Y})$
and $\hat{\mathbf{A}}=(\hat{A}_{x},\hat{A}_{y})$. Consequently, we
can treat only one of the system's equivalent directions. Thus, the Hamiltonian that captures the ion-cavity dynamics is given by
\begin{equation}
\hat{H}_{{\text{ion-cm}}}=-\frac{\hbar^{2}}{2m}\partial_{X}^{2}+\frac{m\Omega^{2}}{2}\hat{X}^{2}+\frac{\textrm{i}g_{0}\hbar}{m}\sqrt{N}\hat{A}_{x}\partial_{X}+\frac{Ng_{0}^{2}}{2m}\hat{A}_{x}^{2}+\hbar\omega\left(\hat{a}_{x}^{\dagger}\hat{a}_{x}+\frac{1}{2}\right).
\end{equation}
Finally, in terms of the matter annihilation operators 
\begin{equation}
\hat{b}_{x}=\sqrt{\frac{m\Omega}{2\hbar}}\left(\hat{X}+\frac{\hbar}{m\Omega}\partial_X\right),
\end{equation}
the CM Hamiltonian attains the form of the well-known HH
\cite{Hopfield, kockum2019ultrastrong} 
\begin{equation}
\hat{H}_{\text{ion-cm}}=\hbar\Omega\left(\hat{b}_{x}^{\dagger}\hat{b}_{x}+\frac{1}{2}\right)-\textrm{i}\sqrt{\frac{\hbar\Omega g_{0}^{2}N}{2m}}\hat{A}_{x}(\hat{b}_{x}^{\dagger}-\hat{b}_{x})+\frac{Ng_{0}^{2}}{2m}\hat{A}_{x}^{2}+\hbar\omega\left(\hat{a}_{x}^{\dagger}\hat{a}_{x}+\frac{1}{2}\right).
\end{equation}

\subsection{Landau Levels \label{sec:APP-Landau}}

Next, we provide the details about the light-matter dynamics of 2degs coupled to the cavity field defined in Eq.(\ref{eq:2degH}). The interaction term between the cavity field and the dynamical momentum of electrons is given by 
\begin{equation}
\hat{\mathbf{A}}\cdot\sum_{i=1}^{N}\textrm{i}\hbar\nabla_{i}+e\hat{\mathbf{A}}_{\mathrm{ext}}(\hat{\mathbf{r}}_{i})=\sqrt{N}\hat{\mathbf{A}}\cdot[\textrm{i}\hbar\nabla_{\mathbf{R}}+e\hat{\mathbf{A}}_{\textrm{ext}}(\hat{\mathbf{R}})].
\label{CM Bfield}
\end{equation}
In terms of new coordinates, the sum of squares of the 
external field is given by
\begin{equation}
\sum_{i=1}^{N}\hat{\mathbf{A}}_{\mathrm{ext}}^{2}(\hat{\mathbf{r}}_{i})=\hat{\mathbf{A}}_{\mathrm{ext}}^{2}(\hat{\mathbf{R}})+N\sum_{j=2}^{N}\hat{\mathbf{A}}_{\mathrm{ext}}^{2}(\hat{\tilde{\mathbf{r}}}_{j})-\left[\sum_{j=2}^{N}\hat{\mathbf{A}}_{\textrm{ext}}(\hat{\tilde{\mathbf{r}}}_{j})\right]^{2}.
\end{equation}
The bilinear term between the magnetic field and the momenta is given by
\begin{equation}
\sum_{i=1}^{N}\hat{\mathbf{A}}_{\mathrm{ext}}(\hat{\mathbf{r}}_{i})\cdot\nabla_{i}=\hat{\mathbf{A}}_{\mathrm{ext}}(\hat{\mathbf{R}})\cdot\nabla_{\mathbf{R}}+\sum_{j=2}^{N}\hat{\mathbf{A}}_{\mathrm{ext}}(\hat{\tilde{\mathbf{r}}}_{j})\cdot\widetilde{\nabla}_{j}.
\end{equation}

Collecting the terms, we find that the Hamiltonian in the new frame is a sum of two terms, $\hat{H}_{\textrm{2deg}}=\hat{H}_{\text{2deg-cm}}+\hat{H}_{\text{2deg-rel}}$, where: (i) the
center of mass term, $\hat{H}_{\textrm{2deg-cm}}$ includes the coupling
to the quantized field $\hat{\mathbf{A}}$, and (ii) the term
$\hat{H}_{\text{2deg-rel}}$ depends on the relative distances, decoupled from $\hat{\mathbf{A}}$. The two terms are given by 
\begin{equation}
\begin{aligned}
\hat{H}_{\text{2deg-cm}} & =\frac{1}{2m}\left(\mathrm{i}\hbar\nabla_{\mathbf{R}}+e\hat{\mathbf{A}}_{\mathrm{ext}}(\hat{\mathbf{R}})+e\sqrt{N}\hat{\mathbf{A}}\right)^{2}+\hbar\omega\left(\hat{a}^{\dagger}\hat{a}+\frac{1}{2}\right),\\
\hat{H}_{\text{2deg-rel}} & =\frac{1}{2m}\sum_{j=2}^{N}\left[\frac{\mathrm{i}\hbar\widetilde{\nabla}_{j}}{\sqrt{N}}+e\sqrt{N}\hat{\mathbf{A}}_{\mathrm{ext}}(\hat{\tilde{\mathbf{r}}}_{j})\right]^{2}-\frac{\hbar^{2}}{2mN}\sum_{j,l=2}^{N}\widetilde{\nabla}_{j}\cdot\widetilde{\nabla}_{l}-\frac{e^{2}}{2m}\left[\sum_{j=2}^{N}\hat{\mathbf{A}}_{\mathrm{ext}}(\hat{\tilde{\mathbf{r}}}_{j})\right]^{2}\\
 & +\sum_{1<l}^{N}W(\sqrt{N}\hat{\tilde{\mathbf{r}}}_{l})+\sum_{2\leq i<l}^{N}W\left(\sqrt{N}(\hat{\tilde{\mathbf{r}}}_{i}-\hat{\tilde{\mathbf{r}}}_{l})\right).
\end{aligned}
\end{equation}

\section{Polariton Branches \label{sec:APP-Frequency-and-period}}

\begin{figure}
\begin{centering}
\includegraphics{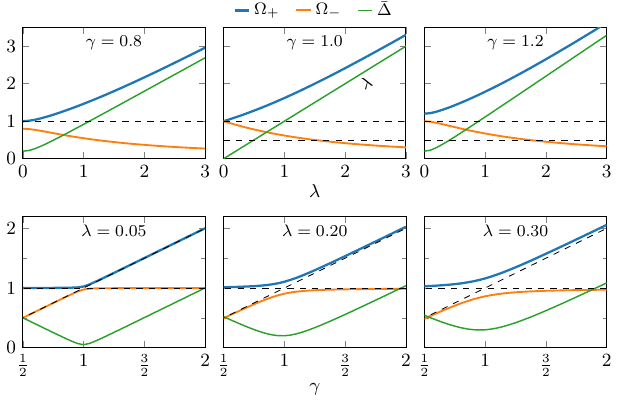} 
\par\end{centering}
\caption{Top panels -- Polariton branches, $\Omega_{\pm}$ and VRS, $\bar{\Delta}\equiv\Omega_{+}-\Omega_{-}$
as a function of the collective coupling constant, $\lambda$. At
resonance, $\bar{\Delta}=\lambda$. Bottom panels -- $\Omega_{\pm}$
and $\bar{\Delta}$ as a function of $\gamma$. The functions' asymmetry
about $\gamma=1$ increases with $\lambda$.}
\label{fig:APP-FIG11} 
\end{figure}

This Appendix presents the polariton frequencies and their dependence
on the collective coupling constant, $\lambda$ and $\gamma\equiv\omega/\Omega$.
Here, $\omega$ is the cavity mode frequency, and $\Omega$ is the
harmonic matter trap frequency. The dimensionless upper/lower polariton frequencies are defined in
Eq.\;\eqref{eq:polariton-freqs} \cite{Vasilis2023}, 
\begin{equation}
\Omega_{\pm}^{2}=\frac{1+\gamma^{2}+\gamma\lambda^{2}}{2}\pm\frac{1}{2}\sqrt{(1+\gamma^{2}+\gamma\lambda^{2})^{2}-4\gamma^{2}},
\label{eq:APP-polariton-branches}
\end{equation}
where the collective coupling constant depends on the number of trapped
particles, $N\geq1$ as $\ensuremath{\lambda\equiv\sqrt{Ng_{0}^{2}/(\pi\epsilon_{0}c\mathcal{A}m\Omega)}}$.
Here, $g_{0}$ is the single particle coupling constant, $\mathcal{A}$
is the area of the resonator mirror, $m$ is the particle mass, $\epsilon_{0}$
is the vacuum permittivity, $c$ is the speed of light. Fig.\;\ref{fig:APP-FIG11}
shows the $\lambda,\gamma$-dependence of $\Omega_{\pm}$ and VRS, $\bar{\Delta}\equiv(\Omega_{+}-\Omega_{-})$. Note that the asymmetry relative to the point $\gamma=1$ increases with $\lambda$.


In the RWA, the terms proportional to
$\lambda\hat{a}^{\dagger}\hat{b}^{\dagger},\lambda\hat{a}\hat{b}$,
and $\lambda^{2}$ in the Hamiltonian [see Eq.\;\eqref{eq:Hcm-dimensionless}]
are omitted, and the polariton branches are given by 
\begin{equation}
\Omega_{\mathrm{RWA},\pm}^{2}=\frac{1+\gamma^{2}+\lambda^{2}/2}{2}\pm\frac{1}{2}\sqrt{(\gamma+1)^{2}[(\gamma-1)^{2}+\lambda^{2}]}.
\label{eq:APP-polariton-branches-RWA}
\end{equation}
Under the RWA, the VRS is symmetric relative to $\gamma=1$. To verify this, it is sufficient to consider $\bar{\Delta}_\text{RWA}^2=(\Omega_{\mathrm{RWA},+}-\Omega_{\mathrm{RWA},-})^2$, which is given by
\begin{equation}
    \bar{\Delta}_\text{RWA}^2 = (\gamma -1)^2+\lambda ^2; \qquad \lambda^2<4\gamma.
\end{equation}
\section{Semiclassical Approach \label{sec:APP-Semiclassical-approach}}

The semiclassical approach relies on solving Hamilton's equations
of motion derived from the classical counterpart of $\hat{\mathcal{H}}$
in Eq.\;\eqref{eq:Hcm-dimensionless} 
\begin{subequations}
\begin{align}
\dot{X} & =P-\lambda q,\quad\dot{P}=-X,\\
\dot{q} & =\gamma p,\quad\dot{p}=-\gamma q+\lambda P-\lambda^{2}q.
\end{align}
\end{subequations}
This system of linear differential equations can be solved
by the Laplace transform method. Laplace transform, denoted by $\mathcal{L}$,
turns the set of differential equations into a system of algebraic
equations since $\mathcal{L}(\dot{f})=\tilde{f}(s)-f(0)$, where
$\tilde{f}(s)$ is the Laplace transform of $f$, $s$ is the Laplace
space variable, and $f(0)$ is the initial value of $f$. Solving
the algebraic system yields, 
\begin{subequations}
\begin{align}
\tilde{X}(s) & =\frac{X(0)[s^{3}+\gamma(\gamma+\lambda^{2})s]+P(0)[s^{2}+\gamma^{2}]}{h(s)}\ -\lambda s\frac{q(0)s+\gamma p(0)}{h(s)},\label{eq:APP-X-Laplace}\\
\tilde{P}(s) & =\frac{[sP(0)-X(0)][s^{2}+\gamma(\gamma+\lambda^{2})]}{h(s)}+\lambda\frac{q(0)s+\gamma p(0)}{h(s)},\\
\tilde{q}(s) & =\gamma\lambda s\frac{P(0)-X(0)}{h(s)}+(s^{2}+1)\frac{q(0)s+\gamma p(0)}{h(s)},\\
\tilde{p}(s) & =\lambda s\frac{P(0)s-X(0)}{h(s)}-\frac{q(0)[\gamma+s^{2}(\gamma+\lambda^{2})]}{h(s)}+\frac{p(0)(s^{3}+s)}{h(s)},
\label{eq:APP-p-Laplace}
\end{align}
\end{subequations}
where $h(s)=s^{4}+s^{2}[\gamma(\gamma+\lambda^{2})+1]+\gamma^{2}$
is the characteristic polynomial of the linear system, and $X(0)$,
$P(0)$, $q(0)$, and $p(0)$ are the initial values. The roots of $h(s)$ are given by $s_{1,2}=\mathrm{i}\Omega_{\pm}$ and $s_{3,4}=-\mathrm{i}\Omega_{\pm}$,
where $\Omega_{\pm}$ are the upper/lower polariton frequencies in
Eq.\;\eqref{eq:APP-polariton-branches}. With those roots, $X(t)$,
$P(t)$, $q(t)$ and $p(t)$ can be found by applying the inverse
Laplace transform, $\mathcal{L}^{-1}$ to Eqs.\;\eqref{eq:APP-X-Laplace}-\eqref{eq:APP-p-Laplace}. The expectation values are obtained semiclassically by averaging the classical
expression for an observable $\hat{O}(t)$, $O(t)=f[X(t),P(t),q(t),p(t)]$
with respect to the initial phase space distribution $\mathcal{P}(0)$
\begin{equation}
\braket{\hat{O}(t)}=\idotsint_{-\infty}^{\infty}O(t)\mathcal{P}(0)\,d\mathbb{V},
\label{eq:APP-phase-space-average}
\end{equation}
where $d\mathbb{V}\equiv dX(0)dP(0)dq(0)dp(0)$. The phase space distribution
is derived from the initial wave function and thus accounts for the
position-momentum uncertainty. For example, the wave function, $\psi_{0}(X,q)\propto\exp[-(X-X_{0})^{2}/(2w^{2})]\exp(-q^{2}/2)$
corresponds to the phase space distribution 
\begin{equation}
\mathcal{P}(0)=\frac{1}{\pi^{2}}e^{-[X(0)-X_{0}]^{2}w^{-2}-P^{2}(0)w^{2}}e^{-q^{2}(0)-p^{2}(0)},
\label{eq:APP-P(0)}
\end{equation}
where $X_{0}$ is the initial displacement of the matter CM quadrature and $w$
is the initial width of the state (for the decoupled CM ground state,
$w=1$).\\

\subsection{Expectation Value of the Matter CM \protect{$X$} Quadrature}

In the case of $\braket{\hat{X}(t)}$, only the first term $\propto X(0)$
[see Eq.\;\eqref{eq:APP-X-Laplace}] contributes to $X(t)$ after
the integration in Eq.\;\eqref{eq:APP-phase-space-average}. Simplification
yields [same as Eq.\;\eqref{eq:<x(t)>-matter}] 
\begin{equation}
\frac{\braket{\hat{X}(t)}}{X_{0}}=\cos\left[\frac{\bar{\Sigma}t}{2}\right]\cos\left[\frac{\bar{\Delta}t}{2}\right]+\beta\sin\left[\frac{\bar{\Sigma}t}{2}\right]\sin\left[\frac{\bar{\Delta}t}{2}\right],
\label{eq:APP-<x(t)>}
\end{equation}
where $\bar{\Sigma}\equiv\Omega_{+}+\Omega_{-}\ensuremath{,}\,\bar{\Delta}\equiv\Omega_{+}-\Omega_{-}$
is the VRS ($\bar{\Sigma}<\bar{\Delta}$), and $\beta=(\Omega_{+}^{2}+\Omega_{-}^{2}-2)/(\bar{\Sigma}\bar{\Delta})$.
Under RWA, 
\begin{equation}
\frac{\braket{\hat{X}(t)}_{\mathrm{RWA}}}{X_{0}}=\cos\left[\frac{\bar{\Sigma}_{\mathrm{RWA}}t}{2}\right]\cos\left[\frac{\bar{\Delta}_{\mathrm{RWA}}t}{2}\right]+\beta_{\mathrm{RWA}}\sin\left[\frac{\bar{\Sigma}_{\mathrm{RWA}}t}{2}\right]\sin\left[\frac{\bar{\Delta}_{\mathrm{RWA}}t}{2}\right],
\label{eq:APP-<x(t)>-RWA}
\end{equation}
where $\bar{\Sigma}_{\mathrm{RWA}}$, $\bar{\Delta}_{\mathrm{RWA}}$,
$\beta_{\mathrm{RWA}}$ are defined in terms of $\Omega_{\text{RWA},\pm}$.\\

\begin{figure}
\begin{centering}
\includegraphics{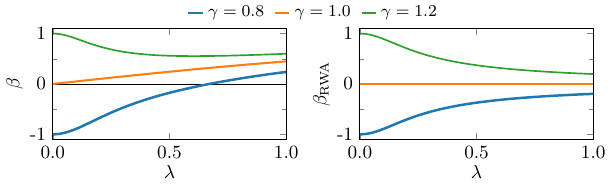} 
\par\end{centering}
\caption{Factors $\beta$ and $\beta_{\mathrm{RWA}}$ in Eqs.\;\eqref{eq:APP-<x(t)>}\;and\;\eqref{eq:APP-<x(t)>-RWA}.
$\beta_{\mathrm{RWA}}$ vanishes at resonance, while $\beta(\gamma=1)$
scales approximately linearly with $\lambda$ (for $\lambda<1$).
Moreover, $|\beta|$ is sensitive to the sign of the relative detuning,
$\gamma-1$, while $|\beta_{\mathrm{RWA}}|$ is not.}
\label{fig:APP-FIG12} 
\end{figure}

\textbf{\emph{$\boldsymbol{\beta}$ factor.}}--Fig.\;\ref{fig:APP-FIG12} compares $\beta$ and $\beta_{\mathrm{RWA}}$.
While at resonance $\beta$ scales approximately linearly with $\lambda$
(for $\lambda<1$), $\beta_{\mathrm{RWA}}(\gamma=1)$ vanishes.
This suggests that the ultra-strong coupling effects in $\braket{X(t)}$,
induced by CRT, are most emphasized at resonance.

\section{Probability Density and Phase Space Distribution \label{sec:APP-Probability-density}}

In this section, we outline the semiclassical derivation of the exact matter
CM probability density, $\mathcal{P}(X,t)$, and discuss its behavior
in Scheme I (for shifted and squeezed matter initial state). Additionally,
we discuss the general form of the matter and light phase space distributions. First, we invert $X[t;X(0)]$ to obtain $X(0)$ as a function of time
and the rest of the initial conditions 
\begin{equation}
X(0)=\frac{X(t)-f_{2}(t)P(0)-f_{3}(t)q(0)-f_{4}(t)p(0)}{f_{1}(t)}.
\end{equation}
The time-dependent functions, $f_{1,\dots,4}$ are given by 
\begin{subequations}
\begin{align}
f_{1}(t) & =\braket{\hat{X}(t)}/X_{0},\\
f_{2}(t) & =\frac{(\Omega_{+}^{2}-\gamma^{2})\sin(\Omega_{+}t)}{\Omega_{+}\bar{\Delta}\bar{\Sigma}}-\frac{(\Omega_{-}^{2}-\gamma^{2})\sin(\Omega_{-}t)}{\Omega_{-}\bar{\Delta}\bar{\Sigma}},\\
f_{3}(t) & =-\lambda\frac{\Omega_{+}\sin(\Omega_{+}t)-\Omega_{-}\sin(\Omega_{-}t)}{\bar{\Delta}\bar{\Sigma}},\\
f_{4}(t) & =\lambda\gamma\frac{\cos(\Omega_{+}t)-\cos(\Omega_{-}t)}{\bar{\Delta}\bar{\Sigma}}.
\end{align}
\end{subequations}
 Substituting the obtained expression for $X(0)$ into the initial
phase space distribution in Eq.\;\eqref{eq:APP-P(0)}, and integrating
out $P(0)$, $q(0)$, and $p(0)$, yields 
\begin{equation}
\mathcal{P}(X,t)=\frac{\exp\left[-\frac{[X-f_{1}X_{0}]^{2}}{w^{2}f_{1}^{2}+f_{2}^{2}/w^{2}+f_{3}^{2}+f_{4}^{2}}\right]}{\sqrt{\pi}\sqrt{w^{2}f_{1}^{2}(t)+f_{2}^{2}/w^{2}+f_{3}^{2}+f_{4}^{2}}}.
\end{equation}
It can be shown that 
\begin{equation}
w^{2}f_{1}^{2}(t)+f_{2}^{2}(t)/w^{2}+f_{3}^{2}(t)+f_{4}^{2}(t)=2\braket{[\Delta\hat{X}(t)]^{2}},
\end{equation}
where $\braket{[\Delta\hat{X}(t)]^{2}}\equiv\braket{\hat{X}^{2}(t)}-\braket{\hat{X}(t)}^{2}$.
Thus, 
\begin{equation}
\mathcal{P}(X,t)=\frac{1}{\sqrt{2\pi\braket{[\Delta\hat{X}(t)]^{2}}}}\exp\left[-\frac{(X-\braket{\hat{X}(t)})^{2}}{2\braket{[\Delta\hat{X}(t)]^{2}}}\right].
\label{eq:APP-P(x,t)}
\end{equation}
\begin{figure}
\begin{centering}
\includegraphics{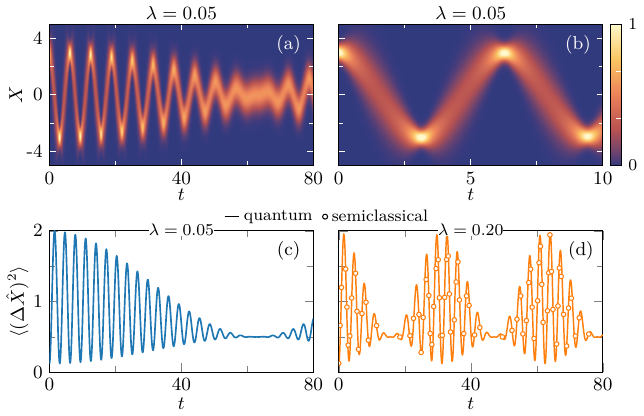} 
\par\end{centering}
\caption{(a,b) Matter CM quadrature probability density, $\mathcal{P}(X,t)$ in Eq.\;\eqref{eq:APP-P(x,t)}.
The wave packet motion combines oscillations and breathing about the
mean value. The total probability is conversed, which causes the bright
spots when the width of the wave packet is minimized. (c,d) Variance
of $\mathcal{P}(X,t)$. Local minima are attained at the turning points
of the wave packet, i.e., at the extrema of $\braket{\hat{X}(t)}$).
$\text{Parameters:\;}\gamma=1,\,X_{0}=3,\,w=0.5$.}
\label{fig:APP-FIG13} 
\end{figure}

Fig.\;\ref{fig:APP-FIG13}(a,b) shows an example of $\mathcal{P}(X,t)$.
The CM wave packet combines oscillatory and breathing motions about
the mean. The wave packet width (variance) increases between the
turning points, as seen in Fig.\;\ref{fig:APP-FIG13}(b). The bright
spots in the density plots appear when the width is minimum, since
the amplitude is maximized at those moments to conserve the total
probability.\\

\textbf{\emph{Phase space distributions.}}--Generally, phase space
distributions of both matter and light subsystems are multivariate
(two-dimensional) Gaussians. A Gaussian function in two dimensions
($X$ and $P$, or $q$ and $p$) depends on six parameters, the coordinates
of the center [$\mu_{X,P}(t)$ or $\mu_{q,p}(t)$], and the elements
of the covariance matrix 
\begin{equation}
\mathbf{\Sigma}(t)=\begin{bmatrix}\sigma_{X}^{2}(t) & \sigma_{XP}(t)\\
\sigma_{XP}(t) & \sigma_{P}^{2}(t)
\end{bmatrix}\quad\text{or}\quad\mathbf{\Sigma}(t)=\begin{bmatrix}\sigma_{q}^{2}(t) & \sigma_{qp}(t)\\
\sigma_{qp}(t) & \sigma_{p}^{2}(t)
\end{bmatrix}.
\end{equation}
For example, the photon phase space distribution is given by 
\begin{align}
\mathcal{P}(q,p,t)=\frac{1}{2\pi\sqrt{\sigma_{q}^{2}\sigma_{p}^{2}-\sigma_{qp}^{2}}}\exp\left[-\frac{\sigma_{p}^{2}(q-\mu_{q})^{2}}{2(\sigma_{q}^{2}\sigma_{p}^{2}-\sigma_{qp}^{2})}\right]\exp\left[-\frac{\sigma_{q}^{2}(p-\mu_{p})^{2}}{2(\sigma_{q}^{2}\sigma_{p}^{2}-\sigma_{qp}^{2})}+\frac{2\sigma_{qp}(q-\mu_{q})(p-\mu_{p})}{2(\sigma_{q}^{2}\sigma_{p}^{2}-\sigma_{qp}^{2})}\right],
\label{eq:APP-P(q,p)}
\end{align}
where $\sigma_{\nu}^{2}\equiv\braket{(\Delta\hat{\nu})^{2}}=\braket{\hat{\nu}^{2}}-\mu_{\nu}^{2}$,
$\sigma_{qp}\equiv\braket{(\Delta\hat{q}\hat{p})}=\braket{\hat{q}\hat{p}}-\mu_{q}\mu_{p}$,
and $\mu_{\nu}\equiv\braket{\hat{\nu}}$ ($\nu=q,p$). Phase space
distribution such as in Eq.\;\eqref{eq:APP-P(q,p)} appear as rotated
ellipses centered around $(\mu_{q},\mu_{p})$. The eigenvectors of the
covariance matrix define the orientation of the ellipses, while the
aspect ratio is proportional to the ratio of the eigenvalues.

Higher-order expectation values related to the matter or light subsystems
can be conveniently obtained by averaging the corresponding classical
expressions with respect to the phase space distributions. The higher-order
averages over a two-dimensional Gaussian distribution are entirely
determined by the distribution's first two moments, namely the means
[e.g., $\mu_{X,P}(t)$ or $\mu_{q,p}(t)$] and the elements of
$\mathbf{\Sigma}$.

\section{Photon Number Expectation Value \label{sec:APP-Photon-number-expectation}}

In this section, we explore the properties of photon number expectation
value in Schemes I and II. In both cases, we consider the RWA, allowing
us to isolate the ultrastrong-coupling effects arising from the counter-rotating
and diamagnetic terms of the Hamiltonian.

\subsection{Scheme I}

Using the phase space distribution in Eq.\;\eqref{eq:APP-P(q,p)}
the expectation value of the photon number $\ensuremath{\hat{n}=\hat{a}^{\dagger}\hat{a}=(\hat{q}^{2}+\hat{p}^{2})/2-1/2}$
can be conveniently found by following the semiclassical approach
outlined in Sec.\;\ref{sec:APP-Semiclassical-approach}. To simplify
the expression, we subtract the relatively small average photon number
when the CM is in the ground state. Thus, the considered quantity
is $\Delta\braket{\hat{n}(t)}\equiv\braket{\hat{n}(t)}-\braket{\hat{n}(t;X_{0}=0,w=1)}$,
which is given by 
\begin{align}
\Delta\braket{\hat{n}(t)} & =\lambda^{2}(w^{-2}-1)\left(\frac{\gamma^{2}\sin^{2}(\bar{\Sigma}t/2)\sin^{2}(\bar{\Delta}t/2)}{\bar{\Sigma}^{2}\bar{\Delta}^{2}}+\frac{[\Omega_{+}\sin(\bar{\Sigma}t/2)\cos(\bar{\Delta}t/2)-(\bar{\Sigma}/2)\sin(\Omega_{-}t)]^{2}}{\bar{\Sigma}^{2}\bar{\Delta}^{2}}\right)\nonumber \\
+\lambda^{2} & (2X_{0}^{2}+w^{2}-1)\left(\gamma^{2}\frac{[\Omega_{-}\sin(\bar{\Sigma}t/2)\cos(\bar{\Delta}t/2)-(\bar{\Sigma}/2)\sin(\Omega_{-}t)]^{2}}{\Omega_{+}^{2}\Omega_{-}^{2}\bar{\Sigma}^{2}\bar{\Delta}^{2}}+\frac{\sin^{2}(\bar{\Sigma}t/2)\sin^{2}(\bar{\Delta}t/2)}{\bar{\Sigma}^{2}\bar{\Delta}^{2}}\right),
\label{eq:delta-n-Scheme-I}
\end{align}
The time-averaged value reads 
\begin{equation}
\Delta\overline{\braket{\hat{n}}}=\lambda^{2}(w^{-2}-1)f_{1}+\lambda^{2}(2X_{0}^{2}+w^{2}-1)f_{2},
\end{equation}
where $f_{1,2}$ [see Eq.\;\eqref{eq:n-time-averaged}] are defined
as 
\begin{subequations}
\begin{align}
f_{1} & =\frac{\Omega_{+}^{2}+\Omega_{-}^{2}+2\gamma^{2}}{8\bar{\Sigma}^{2}\bar{\Delta}^{2}}=\frac{1}{8}\frac{\gamma(3\gamma+\lambda^{2})+1}{\gamma^{2}[(\gamma+\lambda^{2})^{2}-2]+2\gamma\lambda^{2}+1},\\
f_{2} & =\frac{2\Omega_{+}^{2}\Omega_{-}^{2}+\gamma^{2}(\Omega_{+}^{2}+\Omega_{-}^{2})}{8\Omega_{+}^{2}\Omega_{-}^{2}\bar{\Sigma}^{2}\bar{\Delta}^{2}}=\frac{1}{8}\frac{\gamma(\gamma+\lambda^{2})+3}{\gamma^{2}[(\gamma+\lambda^{2})^{2}-2]+2\gamma\lambda^{2}+1}.
\end{align}
\end{subequations}
At resonance, $f_{1,2}$ simplify to $f_{1}=f_{2}=1/(8\lambda^{2})$,
and $\Delta\overline{\braket{\hat{n}}}$ becomes independent of $\lambda$,
\begin{equation}
\Delta\overline{\braket{\hat{n}(\gamma=1)}}=\frac{X_{0}^{2}}{4}+\frac{1}{8}(w-w^{-1})^{2}.
\label{eq:APP-Scheme-I-n-resonance}
\end{equation}
When $w\sim1$, $\Delta\overline{\braket{\hat{n}}}$ can be roughly
estimated by the single term proportional to $X_{0}^{2}$, 
\begin{equation}
\Delta\overline{\braket{\hat{n}}}\approx X_{0}^{2}\cdot(2\lambda^{2}f_{2}).
\label{eq:APP-Scheme-I-n-approx}
\end{equation}
Fig.\;\ref{fig:FIG5}(a,b) shows $\Delta\overline{\braket{\hat{n}}}$ as a function
of $\lambda$ and $\gamma$. In panel (b), the asymmetry about the
point of resonance, $\gamma=1$ increases with $\lambda$. Next, by
comparison with the RWA, we demonstrate that the asymmetry stems
from the ultrastrong coupling and CRT. \\

\textbf{\emph{Rotating Wave Approximation.}}--In the RWA, the Hamiltonian
takes the form 
\begin{equation}
\hat{\mathcal{H}}_\text{RWA}=\left[\hat{b}^{\dagger}\hat{b}+\frac{1}{2}\right]+\gamma\left[\hat{a}^{\dagger}\hat{a}+\frac{1}{2}\right]-\mathrm{i}\frac{\lambda}{2}\hat{b}^{\dagger}\hat{a}+\mathrm{i}\frac{\lambda}{2}\hat{b}\hat{a}^{\dagger}.
\label{eq:APP-RWA-Hamiltonian}
\end{equation}
Expressing $\hat{\mathcal{H}}_\text{RWA}$ in terms
of $\hat{X}$, $\hat{P}$, $\hat{q}$, and $\hat{p}$, and following
the semiclassical approach, we can obtain the photon number expectation
value 
\begin{align}
\braket{\hat{n}(t)}_{\mathrm{RWA}} & =\lambda^{2}\left[2X_{0}^{2}+(w-w^{-1})^{2}\right]\frac{(\gamma+1)^{2}\sin^{2}(\bar{\Sigma}_{\mathrm{RWA}}t/2)\sin^{2}(\bar{\Delta}_{\mathrm{RWA}}t/2)}{4\bar{\Sigma}_{\mathrm{RWA}}^{2}\bar{\Delta}_{\mathrm{RWA}}^{2}}+\lambda^{2}\left[2X_{0}^{2}+(w-w^{-1})^{2}\right]\nonumber \\
\times & \frac{\left[\Omega_{\mathrm{RWA},-}(\gamma\!-\!\lambda^{2}/4\!+\!\Omega_{\mathrm{RWA},+}^{2})\sin(\Omega_{\mathrm{RWA},+}t)-\Omega_{\mathrm{RWA},+}(\gamma\!-\!\lambda^{2}/4\!+\!\Omega_{\mathrm{RWA},-}^{2})\sin(\Omega_{\mathrm{RWA},-}t)\right]^{2}}{16\Omega_{\mathrm{RWA},+}^{2}\Omega_{\mathrm{RWA},-}^{2}\bar{\Sigma}_{\mathrm{RWA}}^{2}\bar{\Delta}_{\mathrm{RWA}}^{2}},\label{eq:APP-delta-n-Scheme-1-RWA}
\end{align}
where $\Omega_{\mathrm{RWA},\pm}$ are defined in Eq.\;\eqref{eq:APP-polariton-branches-RWA}.
Note that in the full system, $\overline{\braket{\hat{n}}}$ differs
from zero for the CM ground state ($X_{0}=0,\,w=1$). That is why
we defined $\Delta\braket{\hat{n}(t)}$ in Eq.\;\eqref{eq:delta-n-Scheme-I}
to simplify the analysis. In contrast, under RWA, $\braket{\hat{n}(t;X_{0}=0,w=1)}\equiv0$
as seen from Eq.\;\eqref{eq:APP-delta-n-Scheme-1-RWA}. The time average of $\braket{\hat{n}(t)}_{\mathrm{RWA}}$ is given
by the simple expression
\begin{equation}
\overline{\braket{\hat{n}}}_{\mathrm{RWA}}=\frac{1}{8}\frac{\lambda^{2}}{(\gamma-1)^{2}+\lambda^{2}}\left[2X_{0}^{2}+(w-w^{-1})^{2}\right].
\end{equation}
At resonance ($\gamma=1$), $\overline{\braket{\hat{n}}}_{\mathrm{RWA}}$
reduces to Eq.\;\eqref{eq:APP-Scheme-I-n-resonance}. Under RWA,
the prefactor of $X_{0}^{2}$ (in this case, it is the overall prefactor)
is a Lorentzian function of $\gamma$ of width $\lambda$ centered
at $\gamma=1$, 
\begin{equation}
2\lambda^{2}f_{\mathrm{RWA}}=\frac{1}{4}\frac{\lambda^{2}}{(\gamma-1)^{2}+\lambda^{2}}.
\label{eq:APP-x0-prefactor-RWA}
\end{equation}

\section{Mandel \protect{$Q$} function \label{sec:APP-Mandel-Q-function}}

The first subsection presents the analytical expression for the \emph{initial}
$Q$ function in the matter subsystem in Scheme I. The following subsections
discuss the impact of the diamagnetic term and the RWA on $Q$.

\subsection{Initial \protect{$Q_\text{mat}$} Function in Scheme I}

To obtain the analytical expression for the initial Mandel $Q$ function
of the matter state, $Q_{\mathrm{mat}}(0)$ in Scheme I (i.e., for
a shifted by $X_{0}$ and squeezed Gaussian), we use the formulas
for $\braket{\hat{n}}$ and $\braket{(\Delta\hat{n})^{2}}$ in Eqs.\;(7.100)\;and\;(7.102)
in \cite{KnightBook2012} 
\begin{align}
\braket{\hat{n}} & =|\alpha|^{2}+\sinh^{2}r,\quad\braket{(\Delta\hat{n})^{2}}=|\alpha|^{2}e^{2r}+2\sinh^{2}r\cosh^{2}r.
\end{align}
For the considered Gaussian initial state, $|\alpha|^{2}=X_{0}^{2}/2$
and $r=-\ln(w)$. Substitution yields 
\begin{align}
Q_{\mathrm{mat}}(0) & \equiv\frac{\braket{(\Delta\hat{n})^{2}}-\braket{\hat{n}}}{\braket{\hat{n}}}=\frac{(w^{2}-1)^{2}(w^{4}+1)+4w^{4}(w^{2}-1)X_{0}^{2}}{4w^{4}X_{0}^{2}+2w^{2}(w^{2}-1)^{2}}.
\end{align}
Expanding in powers of $X_{0}^{-2}$, yields 
\begin{align}
Q_{\mathrm{mat}}(0) & =(w^{2}-1)-\frac{(w^{2}-1)^{2}(w^{4}-2w^{2}-1)}{4w^{4}X_{0}^{2}}+\mathcal{O}(X_{0}^{-4}).
\end{align}
Thus, we obtain the expression used in the main text,
$Q_{\mathrm{mat}}(0)=(w^{2}-1)[1+\mathcal{O}(X_{0}^{-2})]$.

\subsection{Diamagnetic Term and RWA in Scheme I}

\begin{figure}
\begin{centering}
\includegraphics{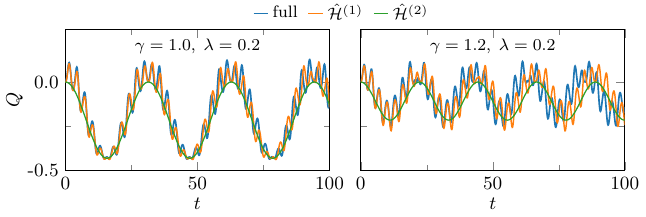} 
\par\end{centering}
\caption{Photon $Q$ function in Scheme I obtained by direct numerical solution
of the Schr\"odinger equation with the HH in Eq.\;\eqref{eq:Hcm-dimensionless}
vs $\hat{\mathcal{H}}^{(1,2)}$ in Eqs.\;\eqref{eq:APP-H_cm-no-A2}\;and\;\eqref{eq:APP-H_cm-RWA}.
Initial matter state parameters: $X_{0}=3$, $w=0.5$.}
\label{fig:APP-FIG14} 
\end{figure}
\begin{figure}
\begin{centering}
\includegraphics{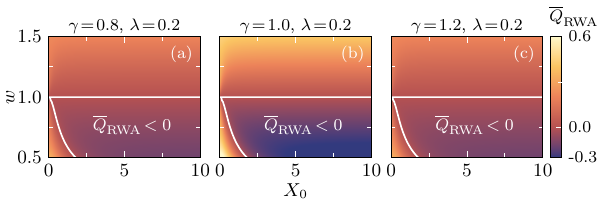} 
\par\end{centering}
\caption{The dependence of photon $\overline{Q}$ on the initial displacement
and width of the matter CM $X$ quadrature under RWA. In contrast to Fig.\;\ref{fig:FIG6}(c-e),
the pictures on either side of the resonance are identical here.}
\label{fig:APP-FIG15}
\end{figure}

For an initially squeezed matter CM state, the photon $Q$ function may attain
negative values even in the absence of the diamagnetic term proportional
to $\lambda^{2}$ i.e., with the Hamiltonian 
\begin{equation}
\hat{\mathcal{H}}^{(1)}=\left[\hat{b}^{\dagger}\hat{b}+\frac{1}{2}\right]+\gamma\left[\hat{a}^{\dagger}\hat{a}+\frac{1}{2}\right]-\mathrm{i}\frac{\lambda}{2}(\hat{b}^{\dagger}-\hat{b})(\hat{a}^{\dagger}+\hat{a}).
\label{eq:APP-H_cm-no-A2}
\end{equation}
Under the RWA, i.e., with the Hamiltonian 
\begin{equation}
\hat{\mathcal{H}}^{(2)}=\left[\hat{b}^{\dagger}\hat{b}+\frac{1}{2}\right]+\gamma\left[\hat{a}^{\dagger}\hat{a}+\frac{1}{2}\right]-\mathrm{i}\frac{\lambda}{2}\hat{b}^{\dagger}\hat{a}+\mathrm{i}\frac{\lambda}{2}\hat{b}\hat{a}^{\dagger},
\label{eq:APP-H_cm-RWA}
\end{equation}
the effective transfer of nonclassical matter state into the cavity
mode is still possible. Fig.\;\ref{fig:APP-FIG14} compares the photon
$Q$ functions obtained with the HH vs $\hat{\mathcal{H}}^{(1,2)}$. To isolate the ultrastrong coupling effect, we reproduce Fig.\;\ref{fig:FIG6}(c-e)
in Fig.\;\ref{fig:APP-FIG15} but under RWA Hamiltonian. $\overline{Q}_{\mathrm{RWA}}$
is insensitive to the sign of the relative detuning $(\gamma-1)$,
i.e., panels (a) and (c) are identical. In contrast, panels (c) and
(e) in Fig.\;\ref{fig:FIG6} exhibit asymmetry induced by the terms
omitted in the RWA. 

\subsection{Diamagnetic Term and RWA in Scheme II}
\begin{figure}
\begin{centering}
\includegraphics{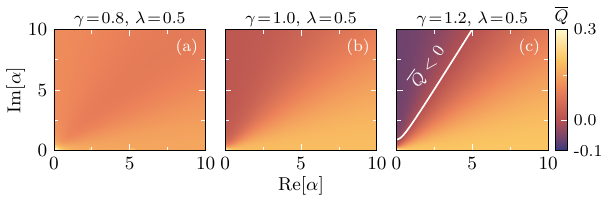} 
\par\end{centering}
\caption{The dependence of photon $\overline{Q}$ on the initial coherent state
of the cavity field. Sub-Poissonian PND is achieved off-resonance,
and the effect is sensitive to the phase of $\alpha$. In the ultra-strong coupling regime here, $\overline{Q}<0$ is completely
absent in the red-shifted cavity. The negativity saturates with an increasing imaginary component of $\alpha$.}
\label{fig:APP-FIG16} 
\end{figure}
\begin{figure}
\begin{centering}
\includegraphics{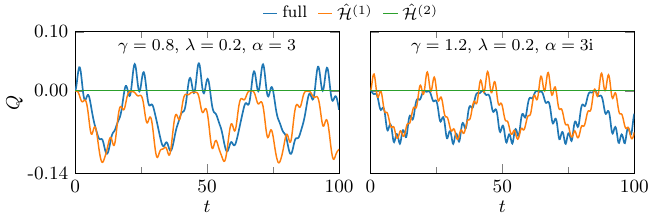} 
\par\end{centering}
\caption{Photon $Q(t)$ function in Scheme II obtained by direct numerical
solution of the Schr\"odinger equation with the HH in
Eq.\;\eqref{eq:Hcm-dimensionless} vs $\hat{\mathcal{H}}^{(1,2)}$
in Eqs.\;\eqref{eq:APP-H_cm-no-A2}~and~\eqref{eq:APP-H_cm-RWA}.}
\label{fig:APP-FIG17} 
\end{figure}
\begin{figure}
\begin{centering}
\includegraphics{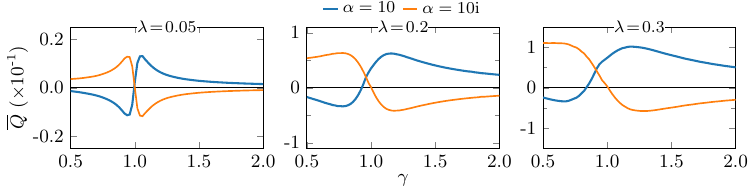} 
\par\end{centering}
\caption{Dependence of the constant component of the photon $Q$ function,
$\overline{Q}$ on $\gamma$ for real/imaginary $\alpha$, defining
the initial coherent state, $\ket{\alpha}$. The chosen values of
$\alpha$ are sufficiently high so that $\overline{Q}$ is saturated
and no longer changes with increasing $\alpha$.}
\label{fig:APP-FIG18} 
\end{figure}
Fig.\;\ref{fig:APP-FIG16} shows additional examples of the constant
component of the photon $Q$ function, $\overline{Q}$ in Scheme II
for a range of initial coherent states of the cavity, $\ket{\alpha}$.
In this example of the ultra-strong coupling regime, the asymmetry
between the red and blue shifted cavities is so pronounced that negative
$\overline{Q}$ is obtained only on one side, for $\gamma=1.2$.

The coherent initial state of the cavity used in Scheme II is not
squeezed, i.e., $Q(t=0)=0$. Therefore, the counter-rotating terms
are essential for generating squeezing and achieving $Q<0$. Fig.\;\ref{fig:APP-FIG17}
shows that under the RWA [i.e., with $\hat{\mathcal{H}}^{(2)}$
in Eq.\;\eqref{eq:APP-H_cm-RWA}], $Q$ remains zero throughout
the motion. In contrast, neglecting the diamagnetic terms [i.e.,
with $\hat{\mathcal{H}}^{(1)}$ in Eq.\;\eqref{eq:APP-H_cm-no-A2}]
produce qualitatively similar results to the full Hamiltonian with
negative (on average) photon $Q$ function. Fig.\;\ref{fig:APP-FIG18} focuses on the asymmetry of $\overline{Q}$
about the point of resonance ($\gamma=1$). Consistent with Fig.\;\ref{fig:FIG9}, the asymmetry between initial coherent states, $\ket{\alpha}$ with
real or imaginary $\alpha$ increases with $\lambda$.

\bibliography{bibliography}

\end{document}